\documentclass[lettersize,journal, twoside]{IEEEtran}
\usepackage{amsmath,amsfonts,amsthm,amssymb}
\usepackage{algorithmic}
\usepackage{algorithm}
\usepackage{array}
\usepackage[caption=false,font=footnotesize,labelfont=rm,textfont=rm]{subfig}
\usepackage{textcomp}
\usepackage{stfloats}
\usepackage{url}
\usepackage{verbatim}
\usepackage{graphicx}
\usepackage{cite}
\usepackage{bm}
\usepackage{color}
\begin{document}

\title{\textcolor{black}{Latent-attention Based Transformer for Near ML Polar Decoding in Short-code Regime}}

\author{Hongzhi Zhu,~ 
        % Wei Xu\IEEEauthorrefmark{1},~\IEEEmembership{Fellow,~IEEE,}~
        % and Xiaohu You\IEEEauthorrefmark{1},~\IEEEmembership{Fellow,~IEEE,}
        Wei Xu,~\IEEEmembership{Fellow,~IEEE,}~
        and Xiaohu You,~\IEEEmembership{Fellow,~IEEE}
        % <-this % stops a space
\thanks{H. Zhu is with the \textcolor{black}{National Mobile Communications Research Laboratory (NCRL), Southeast University, Nanjing, 210096, and the Purple Mountain Laboratories, Nanjing, 210096, China China (email: hz\_zhu@seu.edu.cn).}}%
\thanks{W. Xu and X. You are with the \textcolor{black}{National Mobile Communications Research Laboratory (NCRL),} Southeast University, Nanjing, 210096, China, and the Purple Mountain Laboratories, Nanjing, 210096, China (email: \{wxu, xhyu\}@seu.edu.cn). They are also the corresponding authors of this paper.}
}

% The paper headers
\markboth{}%
{H. Zhu \MakeLowercase{\textit{et al.}}}

\maketitle

\begin{abstract}
\textcolor{black}{Transformer architectures have emerged as promising deep learning (DL) tools for modeling complex sequence-to-sequence interactions in channel decoding. However, current transformer-based decoders for error correction codes (ECCs) demonstrate inferior decoding performance compared to conventional algebraic decoders, especially in short-code regimes. Furthermore, existing DL based channel decoders usually require a dedicated network for each specific code with fixed code length and code rate. This drawback seriously prevents a single model from generalizing across diverse code configurations.} In this work, we propose a novel latent-attention based transformer (LAT) decoder for polar codes that addresses the limitations on performance and generalization through three pivotal innovations. First, we develop a latent-attention mechanism that supersedes the conventional self-attention mechanism. This architectural modification enables independent learning of the Query and Key matrices for code-aware attention computation, decoupling them from the Value matrix to emphasize position-wise decoding interactions while reducing context correlation interference. Second, we devise an advanced training framework incorporating three synergistic components: entropy-aware importance sampling that emphasizes low-probability regions in the signal constellation space, experience reflow that introduces empirical labels to improve characterization of decoding boundaries, and dynamic label smoothing for likelihood-based regularization. Third, we propose a code-aware mask scheme which allows dynamic adaptation for varying code configurations. Numerical evaluations demonstrate that the proposed LAT decoder achieves near maximum-likelihood (ML) performance in terms of both bit error rate (BER) and block error rate (BLER) for short-length polar codes. Furthermore, the architecture exhibits robust generalization capabilities across diverse code rates and code lengths.
\end{abstract}


\section{Introduction}
Channel coding stands as a major role among physical layer technologies in modern communication systems, which fundamentally determines transmission reliability and spectral efficiency\cite{ref0, ref0.1}. The field has undergone remarkable evolution since Shannon's pioneering introduction of the achievable performance bounds in 1948 \cite{ref1}, with seven decades of research dedicated to constructing practical coding schemes that asymptotically approach the Shannon limit\cite{ref2, ref2.1}. The development of modern coding theory has yielded several landmark achievements such as turbo codes\cite{ref3}, low-density parity check (LDPC) codes\cite{ref4}, and polar codes\cite{ref5}. All these codes contribute uniquely to addvancing communication systems through enhanced error-correction capabilities. 

Polar codes, in particular, represent a theoretical breakthrough as the first class of codes that provably achieve the Shannon limit under successive cancellation (SC) decoding for infinite code length\cite{ref6}. This mathematical guarantee has positioned polar codes as strong candidates for next-generation communication standards. However, polar codes face critical challenges in finite-length regimes, where the SC decoder exhibits significant performance degradation compared to the optimal maximum-likelihood (ML) decoder\cite{ref7}. This reliability-complexity trade-off has sparked intensive research efforts to develop novel decoding architectures that reconcile near ML performance of polar codes with polynomial-time complexity, particularly for short-code applications in ultra-reliable low-latency communication (URLLC) and massive machine type communication (mMTC) scenarios\cite{ref8, ref9}.

Recently, the development of deep learning (DL) has catalyzed a paradigm shift in channel coding research by introducing data-driven methodologies that complement conventional algebraic approaches\cite{ref9.1}. Through hierarchical feature extraction from training codewords, DL-based decoders have demonstrated unprecedented capability in resolving mathematically intractable optimization problems\cite{ref10}. This technological leap has been further amplified by specialized DL processors, e.g., Google TPU v4 and NVIDIA A100 Tensor Core, which deliver noticeable energy efficiency for DL decoding tasks \cite{ref11, ref12}, making practical implementation feasible in next-generation communication systems.

In this paper, we propose a novel latent-attention based transformer (LAT) decoder for polar codes that synergizes polar code algebra with neural attention dynamics. The LAT decoder achieves near optimal ML decoding performance in short-code regime and \textcolor{black}{preserves error-correction stability across diverse code lengths and code rates.}

\subsection{Algebraic Decoders for Polar Codes}
SC decoder achieves theoretical optimality for polar codes under infinite blocklength assumptions\cite{ref5}. However, its implementation with finite code lengths in practice reveals fundamental limitations due to severe error propagation induced by the core greedy tree-search mechanism. To mitigate this critical issue, successive cancellation list (SCL) decoder \cite{ref12.1, ref12.2} and successive cancellation stack (SCS) decoder \cite{ref12.3} were proposed as enhancements of the SC decoder. By using a list (or a stack), several temporary sub-optimal decoding paths are maintained concurrently for global optimization. Moreover, through concatenation of cyclic redundancy check (CRC) with SCL decoding, the distance properties of polar codes are further improved \cite{ref7}. While CRC-aided SCL decoder achieves markedly improvement in decoding performance, it imposes considerable decoding complexity and substantial memory overhead, particularly for large decoding lists.  

\subsection{\textcolor{black}{DL-based Decoders for Polar Codes}}
Exploration of deep learning (DL) for channel decoding commenced with multilayer perceptron (MLP) architectures employing fully-connected feedforward neural networks \cite{ref12.5}. While these models demonstrated near ML performance for trivial small code length $N$ (e.g., $N<16$),  MLP architectures suffer exponential training complexity with growing code length and their fixed layer dimensions prevent generalization across varying code lengths. To address these limitations, advanced DL models such as convolutional neural networks (CNN) and recurrent neural networks (RNN) were used to achieve better complexity-performance tradeoffs \cite{ref12.6, ref12.7}. However, since these models mainly focused on local and temporal dependencies, they remain suboptimal for capturing global codeword relationships in polar codes.   

Within the DL ecosystem, transformer architectures have recently emerged as a potent solution for sequence-to-sequence channel decoding. Their core innovation lies in the attention mechanism that dynamically computes pairwise symbol dependencies across the entire received sequence\cite{ref13}. Capitalizing on key advantages such as global feature extraction and parallel computation of all bit decisions, various transformer-based decoders have been proposed for model-free channel decoding. In particular, the error correction code transformer (ECCT) \cite{ref14} utilizes the magnitude of the channel observations and the syndrome to denoise the receive signal through self-attention, and the cross-attention message-passing transformer (CrossMPT) \cite{ref15} separately processes the magnitude and syndrome through dual cross-attention layers in an iterative manner for performance enhancement. While demonstrating compatibility with polar codes, these emerging decoders exhibit substantial reliability deficiencies compared to conventional methods in short-blocklength regimes \cite{ref16}. \textcolor{black}{Moreover, in most cases each single implementation of current transformer-based decoders is inherently bound to a specific code with fixed code length and code rate. This architectural constraint demands complete retraining for varying code configurations.} Accordingly, the generalization capability is severely prevented and thereby a critical impediment is constituted to the practical deployment of these configuration-fixed decoders.  

Further development of unique transformer-based polar decoder requires taking the specific code structure into account. The core self-attention mechanism of transformers, originally designed for natural language processing (NLP) applications, disrupts the inherent polarization structure due to dynamic computation of context correlation. This mismatch brings interference during training and causes serious performance degradation. Consequently, fundamental redesign of attention mechanisms is imperative for polar code decoding optimization.

\subsection{Contributions}
In this paper, we propose a novel LAT decoder for polar codes that successfully synergizes the polar code algebra with neural attention mechanisms. The major contributions of our work are summarized as follows:
\begin{itemize}
\item{We propose a latent-attention mechanism for polar decoding, which explicitly incorporates frozen bit constraints and bit position information into dependency learning, replacing conventional context correlation in self-attention with code-aware attention weights. This mechanism allows the LAT decoder to achieve near ML performance in short-code regime with practical complexity.}
\item{We also propose an efficient training framework, including entropy-aware importance sampling that emphasizes low-probability regions in the signal constellation space, experience reflow to improve characterization of decoding
boundaries, and dynamic label smoothing for likelihood-based regularization. These setups bring significant improvement to the training and maintain compatibility with existing DL-based decoders.}
\item{\textcolor{black}{By developing a code-aware mask scheme that dynamically adapts to varying code configurations, the proposed LAT decoder achieves stable generalization capabilities across diverse code rates and code lengths.} Numerical results demonstrate that LAT decoder closely approaches the bit error rate (BER) and block error rate (BLER) performances of the ML decoder for short polar codes, while the code-aware mask scheme achieves significant improvement on cross-configuration generalization.}
\end{itemize}

\subsection{Paper Outline}
The rest of this paper is organized as follows. In Section \uppercase\expandafter{\romannumeral2}, we present the system model and describe the polar coding mechanism. Section \uppercase\expandafter{\romannumeral3} introduces the structure and mechanism of the proposed LAT decoder. Section \uppercase\expandafter{\romannumeral4} presents further analyses on the mathematical foundation and generalization capabilities of the proposed LAT decoder. Training framework and implementation details of the LAT decoder are elaborated in Section \uppercase\expandafter{\romannumeral5}. The decoding performance of the proposed LAT decoder is evaluated with comparisons in Section \uppercase\expandafter{\romannumeral6}. Conclusions are drawn in Section \uppercase\expandafter{\romannumeral7}.

\subsection{Notations}
In this paper, we let $[n]\triangleq\{1, 2, ..., n\}$ denote the fundamental index set. For a given set $A$, the uniform distribution over $A$ is denoted as ${\rm{Unif}}(A)$. The standard Euclidean inner product between two vectors is denoted as $\langle\cdot,\cdot\rangle$. Multivariate Gaussian distributions are parameterized as $\mathcal{N}(\bm{\mu},\mathbf{\Sigma})$ with mean vector of $\bm{\mu}$ and covariance matrix of $\mathbf{\Sigma}$. Given an $N$-dimensional vector $\bm{x}=(x_1, \dots, x_N)$, we use ${\bm{x}}[i:j]$ to denote a new vector $(x_i, \dots, x_j)$. Given an event $E$, ${\mathbb{I}}(E)$ is the indicator on whether this event happens, and ${\mathbb{P}(E)}$ denotes its probability. Given two vectors $\bm{u}$ and $\bm{v}$, $[\bm{u}, \bm{v}]$ denotes the concatenation of them.   

% and the number of elements inside $A$ is denotes as $\vert A \vert$

\section{Prerequisites}
In this section, we introduce the canonical setting of channel coding and present the polar coding mechanism.

\subsection{System Model}
\begin{figure}[!t]
\centering
\includegraphics[width=3.4 in]{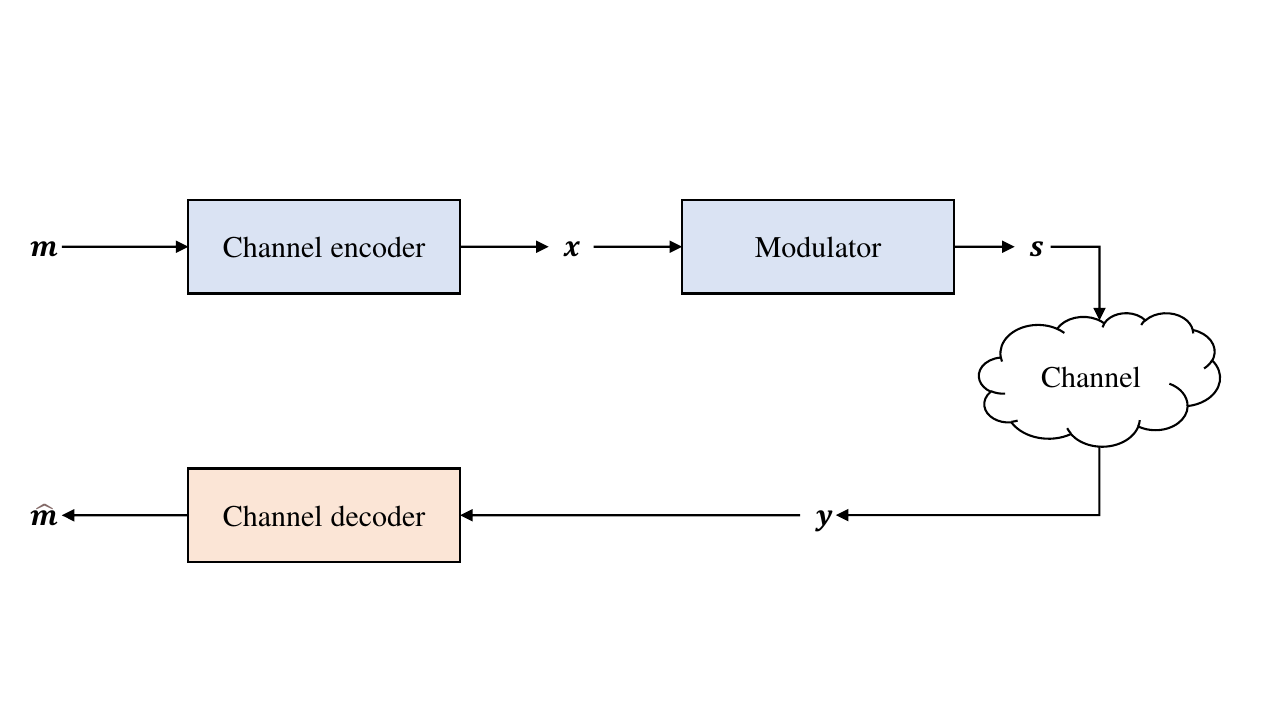}
\caption{An end-to-end system model. The message bit sequence $\bm{m}$ is encoded into $\bm{x}$ and modulated into $\bm{s}$ at the transmitter. $\bm{s}$ is then passed through a channel and is received as $\bm{y}$. At the receiver $\bm{y}$ is decoded to $\bm{\hat{m}}$ as an estimate of $\bm{m}$.} 
\label{fig_system}
\end{figure}

We consider a polar-coded end-to-end communication system as shown in Fig. \ref{fig_system}. In this figure, the message sequence $\bm{m} \in \{0, 1\}^{k}$ of $k$ bits is encoded to a codeword $\bm{x}\in {C}_{\rm{code}}\subset\{0, 1\}^{N}$ of $N$ bits, where ${C}_{\rm{code}}$ denotes the polar code space with code length $N$ and code rate $R=k/N$. $\bm{x}$ is then modulated to $\bm{s}\in \{\pm1\}^{N}$. Following \cite{ref5}, we employ binary phase shift keying (BPSK) modulation, where each code bit $x_i$ is mapped to a BPSK symbol $s_i =1-2{x_i}$. The channel is considered to be additive white Gaussian noise (AWGN) channel. After passing through the channel, a noisy version of $\bm{s}$ is received as $\bm{y}=\bm{s}+\bm{n}$, where $\bm{n}\in \mathbb{R}^{N}\sim\mathcal{N}({\bm{0}}, {\rm{N_0}}{\mathbf{I}}_{N})$ is the noise, ${\rm{N_0}}$ is the channel noise variance, and ${\mathbf{I}}_{N}$ is $N$-dimensional identity matrix. For the proposed LAT decoder, $\bm{y}$ is decoded to $\bm{\hat{m}}\in \{0, 1\}^{k}$. The decoding performance is measured using standard error metrics such as BER and BLER.

\subsection{Polar Coding \label{polar_code}}
We constrain the code length to $N=2^{n}$ where $n\in\{1, 2, 3, ...\}$ since the fundamental building block of polar codes is a $2\times2$ binary polarization kernel defined as 
\begin{equation}
\label{polar_kernel}
\mathbf{F}\triangleq\begin{bmatrix}
1 & 0 \\
1 & 1
\end{bmatrix} \text{.}
\end{equation}
In polarization theory, a length-$N$ bit sequence is considered as $N$ independent channel uses that undergo channel combining and splitting operations during encoding. This process induces polarization in reliabilities of the synthesized channels, quantified via log-likelihood ratio (LLR) analysis. For AWGN channels, Gaussian approximation (GA) is widely adopted as an efficient computational framework for estimating the mean LLRs of these synthesized channels\cite{ref17}. Let $A_{k}\subset[N]$ denote the index set of the $k$ most reliable subchannels, with $A_k^{\rm{c}}$ representing the complement. The polar encoding process embeds $\bm{m}$ into an $N$-dimensional vector $\bm{u}$, where $\bm{u}_{A_{k}}=\bm{m}$ and $\bm{u}_{A_k^{\rm{c}}}$ is set to fixed bit values (typically zeros). These fixed bits, defined as the frozen bits, facilitate error correction as a priori knowledge to both encoder and decoder \cite{ref6}. 

Subsequently, the codeword $\bm{x}$ is generated from $\bm{u}$ through
\begin{equation}
\label{polar_encode}
\bm{x} = {\bm{u}}{\mathbf{G}}_{N} \text{,}
\end{equation}
where the generator matrix $\mathbf{G}_{N}$ follows recursive construction:
\begin{equation}
\mathbf{G}_{N} =
\begin{cases}
\label{polar_encode2}
\mathbf{F}, & \text{$N=2$}\\
\mathbf{B}_{N}(\mathbf{F} \otimes \mathbf{G}_{N/2}), & \text{$N>2$,}
\end{cases}
\end{equation}
where $\mathbf{B}_{N}$ represents a reverse shuffle matrix, and $\otimes$ denotes the Kronecker product. Since $\mathbf{B}_{N}$ only affects index shuffling without any impact on the decoding performance\cite{ref5}, we can safely simplify (\ref{polar_encode}) to 
\begin{equation}
\label{polar_encode3}
\bm{x} = {\bm{u}}{\mathbf{F}}^{\otimes{n}} \text{,}
\end{equation}
where ${\mathbf{F}}^{\otimes{n}}$ represents the $n$-th Kronecker power of ${\mathbf{F}}$. 

\begin{figure}[!t]
\centering
\includegraphics[width=3.4 in]{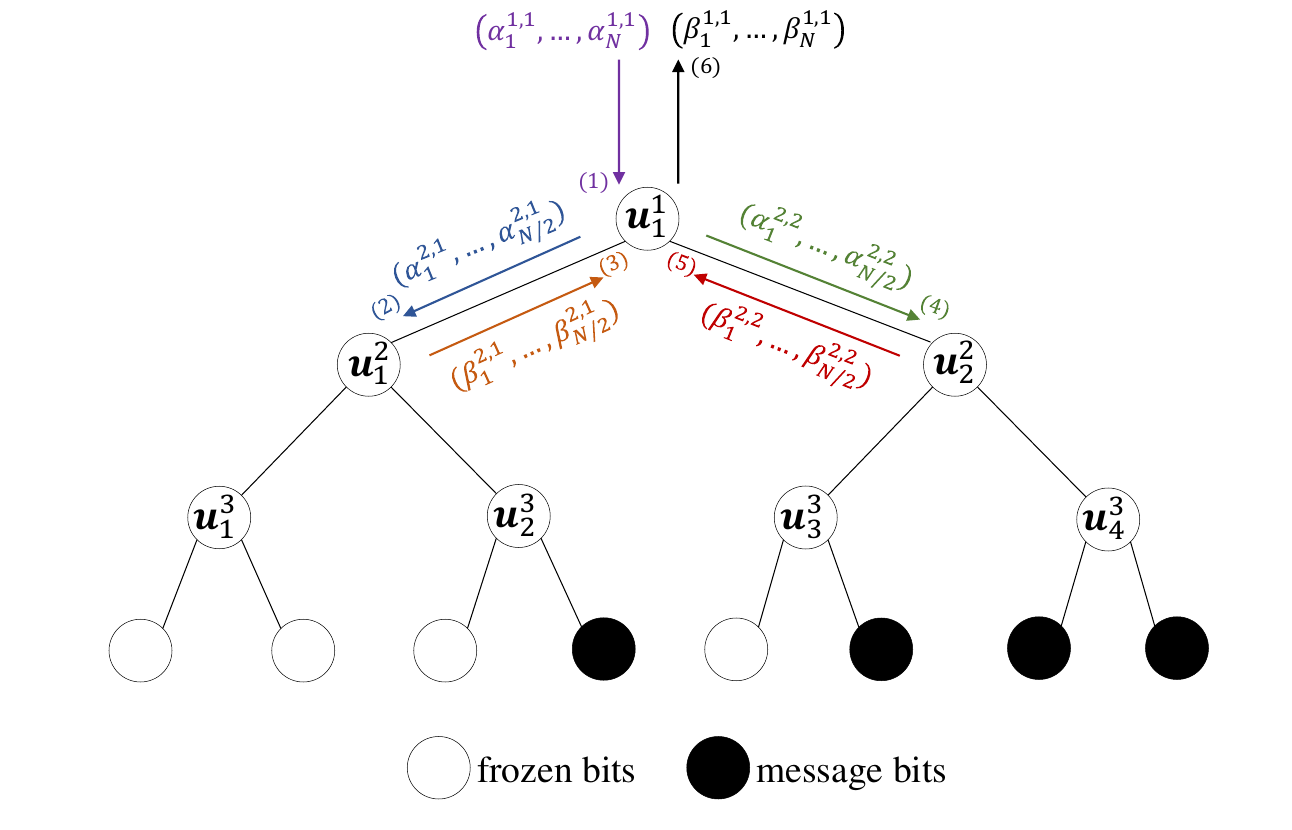}
\caption{Polar code tree. Each node $\bm{u}^{i}_{j}$ contains an LLR vector $\bm{\alpha}^{i,j}$ and a bit value vector $\bm{\beta}^{i,j}$. During SC decoding, each node implements six sequential operations. (1) Input LLR reception from parent. (2) Left-child LLR estimation. (3) Left-child bit decision. (4) Right-child LLR estimation. (5) Right-child bit decision. (6) Self bit decision and feedback to parent. } 
\label{fig_tree}
\end{figure}

The code structure corresponds to a code tree as shown in Fig. \ref{fig_tree}. Let $i\in[n+1]$ denote the layer depth and $j\in[2^{i-1}]$ denote the node index. Each node $\bm{u}^{i}_{j}$ contains an LLR vector denoted as $\bm{\alpha}^{i,j}\in\mathbb{R}^{N/2^{i-1}}$ and a bit value vector denoted as $\bm{\beta}^{i,j}\in\{0,1\}^{N/2^{i-1}}$. The SC decoding implements six nodal operations sequentially while traversing the tree: (1) Input LLR reception from parent. (2) Left-child LLR estimation. (3) Left-child bit decision. (4) Right-child LLR estimation. (5) Right-child bit decision. (6) Self bit decision and feedback to parent.  

As initialization, root node LLRs are derived from channel observations:
\begin{equation}
\label{polar_tree_root}
{\alpha}^{1,1}_{i}={\rm{ln}}{\frac{\mathbb{P}(y_i |0)}{\mathbb{P}(y_i |1)}} \text{,}
\end{equation}
while non-root nodes compute child LLRs via:
\begin{equation}
\begin{cases}
\label{polar_tree_alpha}
{\bm{\alpha}}^{i+1,2j-1}\!=\!{\rm{F}}({\bm{\alpha}}^{i,j}[1:\frac{N}{2^i}], {\bm{\alpha}}^{i,j}[\frac{N}{2^i}+1: \frac{N}{2^{i-1}}]) \\
{\bm{\alpha}}^{i+1,2j}\!=\!{\rm{G}}({\bm{\beta}^{i+1, 2j-1}},{\bm{\alpha}}^{i,j}[1:\frac{N}{2^i}], {\bm{\alpha}}^{i,j}[\frac{N}{2^i}+1: \frac{N}{2^{i-1}}]) \text{,}
\end{cases}
\end{equation}
where the constituent functions are formulates as \cite{ref5}:
\begin{equation}
\label{polar_tree_f}
{\rm{F}}(\bm{x},\bm{y})={{\rm{sign}}(\bm{x}){\rm{sign}}(\bm{y}){\rm{min}}\{|\bm{x}|,|\bm{y}|\}} \text{,}
\end{equation}
\begin{equation}
\label{polar_tree_G}
{\rm{G}}(\bm{u}, \bm{x}, \bm{y})=(1-2\bm{u})\bm{x}+\bm{y} \text{.}
\end{equation}

At the leaf nodes, whose LLRs are scalars, their bit values are decided by
\begin{equation}
\bm{\beta}^{n+1,j}=
\begin{cases}
\label{polar_tree_leaf}
0, & \text{$j\in A_k^{\rm{c}}$}\\
{\mathbb{I}}({\alpha}^{n+1,j}<0), & \text{$j\in A_k$,}
\end{cases}
\end{equation}
while non-leaf nodes decide their bit values through  
\begin{equation}
\label{polar_tree_beta}
\bm{\beta}^{i,j}=[\bm{\beta}^{i+1,2j-1}\oplus\bm{\beta}^{i+1, 2j}, \bm{\beta}^{i+1, 2j}] \text{,}
\end{equation}   
where $\oplus$ denotes the XOR operation. While SC decoders asymptotically reach the Shannon capacity under infinite code length through greedy tree-search strategies, they exhibit significant performance degradation in practical finite-length regimes \cite{ref12.1}. Although the code tree framework supports enhanced algebraic decoders as well (including SCL, SCS, and CRC-aided SCL decoders), these methods incur prohibitive complexity scaling due to redundant code tree traversals \cite{ref7}. To address this inherent efficiency-reliability trade-off, we propose a DL-based LAT decoder that synergizes the polar code structure with neural attention architectures. Through the core latent-attention mechanism and code-aware masking scheme, the proposed LAT decoder enables parallel bit decision and stable generalization capabilities.

\begin{figure*}[!t]
\centering
\includegraphics[width=6 in]{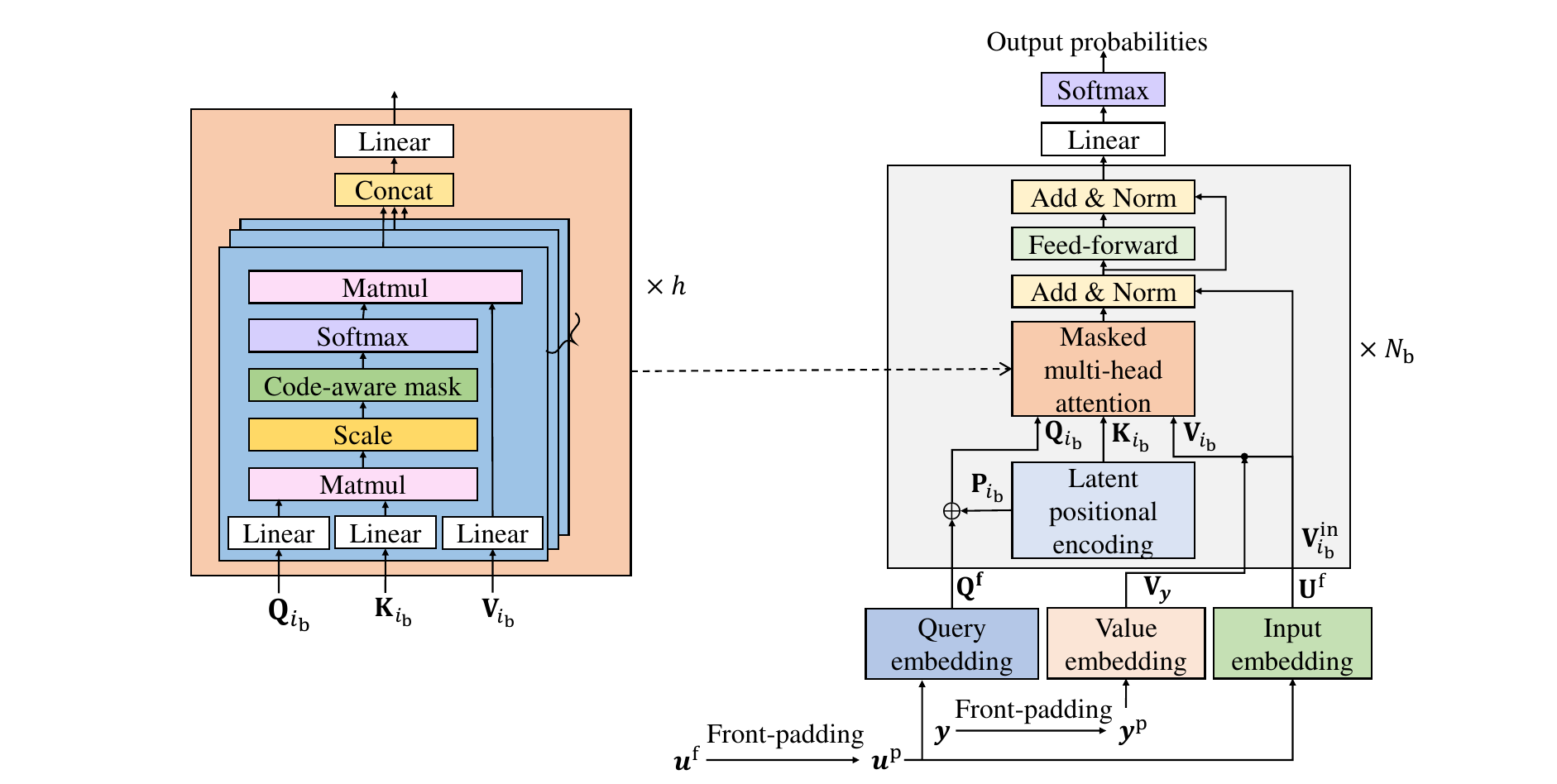}
    \caption{The LAT decoder architecture. Each decoding layer incorporates an LPE that simultaneously generates a latent positional Query encoding ${\mathbf{P}}_{i_{\rm{b}}}$ and a layer-adaptive Key matrix ${\mathbf{K}}_{i_{\rm{b}}}$ for computing hierarchical attention weights. The Query matrix ${\mathbf{Q}}_{i_{\rm{b}}}$ is obtained by summation of ${\mathbf{P}}_{i_{\rm{b}}}$ and a Query embedding ${\mathbf{Q}^{\rm{f}}}$ which carries prior knowledge of frozen bits. The Value matrix $\mathbf{V}_{i_{\rm{b}}}$ is a concatenation of Value embedding $\mathbf{V}_{\bm{y}}$ and the input embedding $\mathbf{U}^{\rm{f}}$. $\mathbf{V}^{\rm{in}}_{i_{\rm{b}}}$ is initialized by input embedding $\mathbf{U}^{\rm{f}}$, and then undergoes progressive refinement through layer-wise propagation.} 
\label{fig_lat}
\end{figure*}

\section{Latent-attention Based Transformer Decoder \label{LAT}}
In this section, we formulate the mechanism of the LAT decoder, including the model architecture and the latent-attention mechanism.

Conventional transformer employs an encoder-decoder architecture to separately extract the features of the duel input sequences. The encoder processes the target sequence while the decoder processes the source sequence and then computes the cross-attention between the two sequences \cite{ref13}. However, the relevance between the channel observation and the frozen prior is independent to the decoding interactions due to the inherent polarization mechanism. Therefore, the proposed LAT decoder adopts a decoder-only transformer architecture where the frozen bits and received signals are concatenated as an integrated attention target. 

As illustrated in Fig. \ref{fig_lat}, the LAT decoder comprises $N_{\rm{b}}$ identical decoding layers. Each decoding layer integrates a latent-attention operation and a fully-connected feed-forward layer, both rounded by residual connections and followed by layer normalization. The model incorporates an $N$-dimensional vector $\bm{u}^{\rm{f}}$ to characterize prior frozen bit knowledge, where
\begin{equation}
\begin{cases}
\label{lat_ini}
\bm{u}^{\rm{f}}_{A_{k}}=0 \\
\bm{u}^{\rm{f}}_{A_k^{\rm{c}}} = 2\bm{u}_{A_k^{\rm{c}}}  - 1 \text{.}
\end{cases}
\end{equation}
To facilitate generalization across diverse code lengths and rates for input codes shorter than the maximum network size, say $N_{\rm{max}}$, a front-padding scheme is employed to expand $\bm{u}^{\rm{f}}$ and $\bm{y}$ into $\bm{u}^{\rm{p}}\in\{0, \pm1\}^{N_{\rm{max}}}$ and $\bm{y}^{\rm{p}}\in \mathbb{R}^{N_{\rm{max}}}$ to fit the network dimension. By extending the code tree in Section \ref{polar_code} into a padded code tree based on the front-padding mechanism, the structural properties of polarization are perfectly preserved due to the inherent absolute bit indices across varying code lengths. $\bm{u}^{\rm{p}}$ then undergoes an input embedding layer into ${\mathbf{U}^{\rm{f}}}$ as an initialization of the recursive decoding layer input $\mathbf{V}^{\rm{in}}_{i_{\rm{b}}}$, progressively updated through layer-wise propagation. Synchronously, through a Value embedding layer and a Query embedding layer, both $\bm{y}^{\rm{p}}$ and $\bm{u}^{\rm{p}}$ are respectively encoded into $\mathbf{V}_{\bm{y}}$ and ${\mathbf{Q}^{\rm{f}}}$, which serve as inherent parameters across all the decoding layers for computation of attention patterns. The output module contains a linear projection and a softmax layer that outputs bitwise probabilities over $\bm{m}$. Through hard decision on the output probabilities, $\bm{\hat{m}}$ is obtained as a binary estimate of $\bm{m}$. \textcolor{black}{The proposed LAT decoder architecture achieves efficient message bit recovery by synergistically leveraging the structured prior frozen bit knowledge and layer-wise positional dependencies of polar codes through the latent-attention mechanism. Furthermore, the front-padding strategy and code-aware masking scheme enables seamless adaptation to varying code configurations. This architectural flexibility endows the LAT decoder with superior generalization capabilities compared to conventional DL decoders which are constrained to fixed code specifications.}

\subsection{Front-padding}
\begin{figure*}[t]
\centering
\subfloat[Post-padding.]{
    \centering
    \begin{minipage}[t]{3.4in}
        \centering       
        \includegraphics[width=\linewidth]{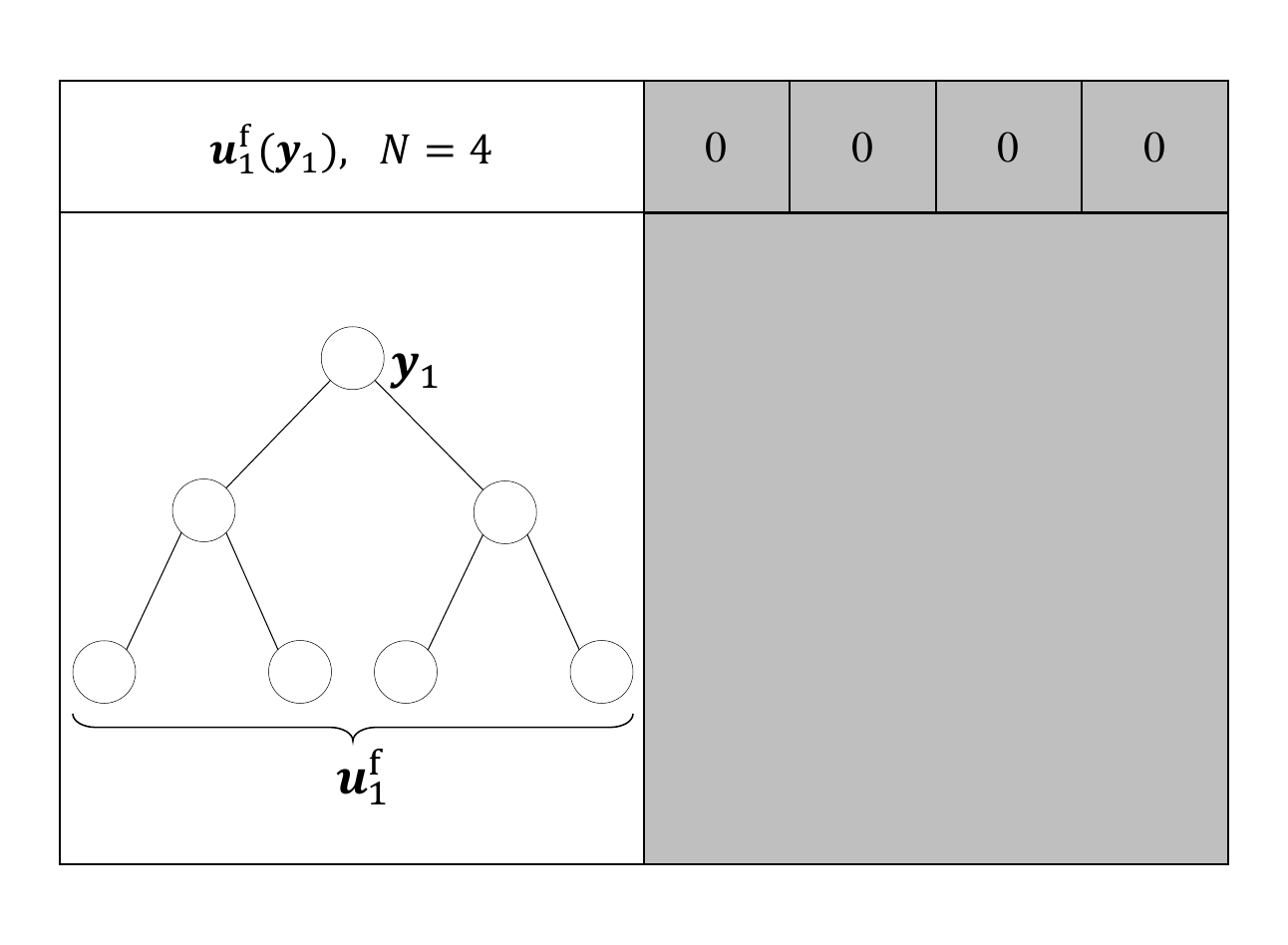}
        \end{minipage}
}%
\subfloat[Front-padding.]{
    \centering
    \begin{minipage}[t]{3.4in}
        \centering       
        \includegraphics[width=\linewidth]{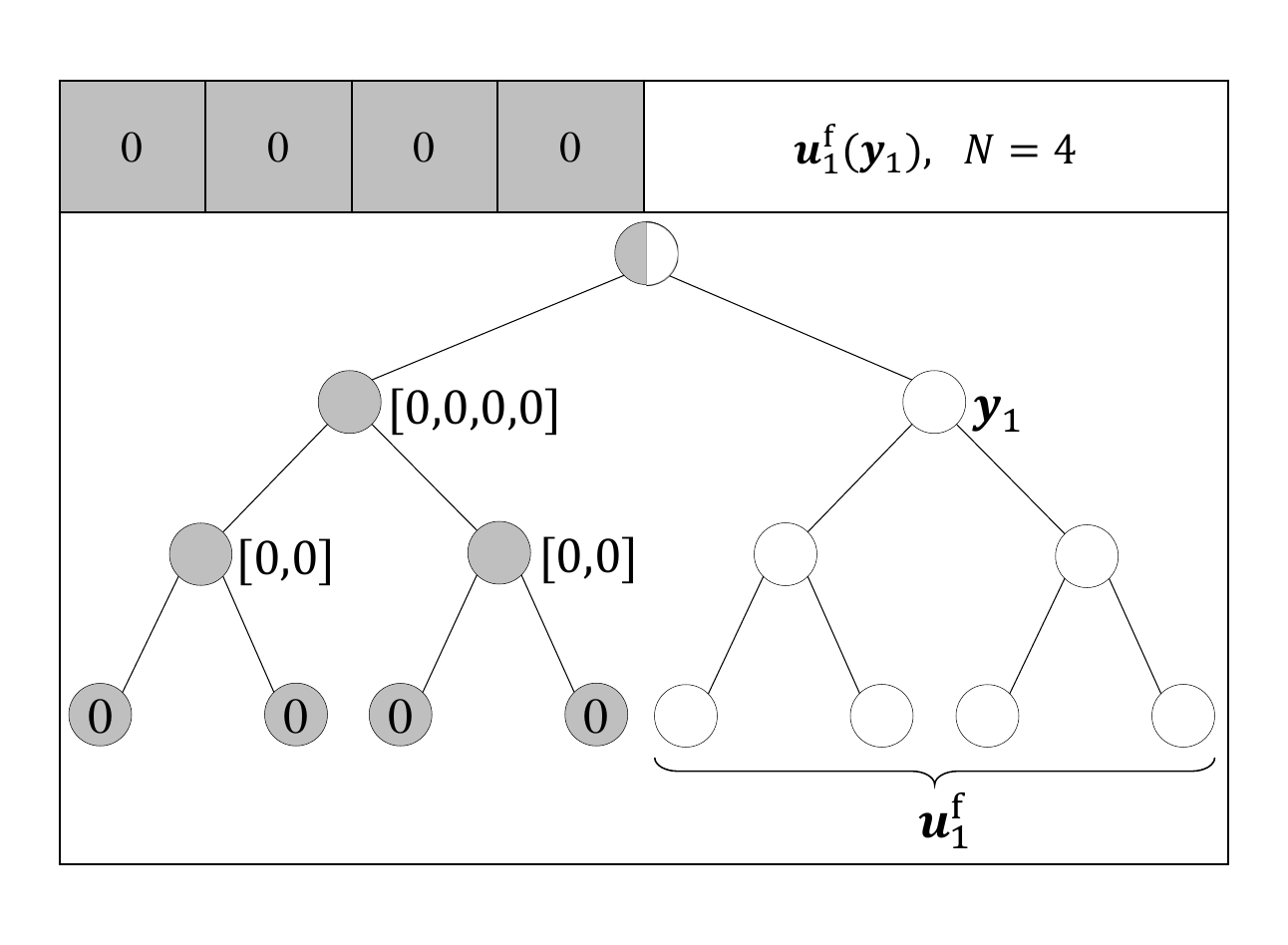}
    \end{minipage}
}%   
\caption{\textcolor{black}{(a) Post-padding and (b) Front-padding for $N=4$ and $N_{\rm{max}}=8$ with the corresponding code trees, where the gray boxes represent padding zeros. Through front-padding, the absolute bit indices are preserved across varying code lengths, which is consistent with their inherent node positions in the padded code tree structures, where gray nodes represent the front-padding zeros and white nodes represent the valid codeword.}}
\label{fig_padding}
% \vspace{-0.5cm}
\end{figure*}

To enable generalization for variable code configurations in practice, we develop a novel prefix padding scheme. Diverging from conventional transformer architectures that adopt post-padding (appending zeros at sequence tails), the proposed front-padding strategy prepends zeros to short codewords, which preserves positional integrity of valid codeword bits by maintaining their absolute positions. Mathematically, for codes with $N < N_{\rm{max}}$, $\bm{u}^{{\rm{p}}}$ and $\bm{y}^{\rm{p}}$ are obtained according to 
\begin{equation}
u^{{\rm{p}}}_{i}=
\begin{cases}
\label{padding_u}
0, & i\in[N_{\rm{pad}}]\\
u^{\rm{f}}_{i-N_{\rm{pad}}}, & i\in[N_{\rm{pad}}+1, N_{\rm{max}}] \text{,}
\end{cases}
\end{equation}
\begin{equation}
y^{{\rm{p}}}_{i}=
\begin{cases}
\label{padding_y}
0, & i\in[N_{\rm{pad}}]\\
y_{i-N_{\rm{pad}}}, & i\in[N_{\rm{pad}}+1, N_{\rm{max}}] \text{,}
\end{cases}
\end{equation}
where $N_{\rm{pad}}=N_{\rm{max}}-N$ represents the padding length. Notably, by extending the binary node value $\bm{\beta}$ of the polar code tree in Section \ref{polar_code} through an affine transformation $\bm{\beta}_{\rm{p}}=2\bm{\beta}-1$ and modifying the non-leaf bit value decision in (\ref{polar_tree_beta}) by
\begin{equation}
\label{polar_tree_beta_new}
\bm{\beta}^{i,j}_{\rm{p}}=[-\bm{\beta}^{i+1,2j-1}_{\rm{p}}\bm{\beta}^{i+1, 2j}_{\rm{p}}, \bm{\beta}^{i+1, 2j}_{\rm{p}}] \text{,}
\end{equation}
the polar code tree is completely adapted to the padded inputs as a padded code tree since both the left-padded zeros and right-aligned valid codewords are perfectly maintained throughout the modified tree without cross-position interference. The diagrams of conventional post-padding and the proposed front-padding under $N=4$ and $N_{\rm{max}}=8$ are respectively demonstrated in Fig. \ref{fig_padding}(a) and Fig. \ref{fig_padding}(b) with their corresponding code trees. Through the front-padding, the absolute bit indices are preserved across varying code lengths, which is consistent with their inherent node positions in the padded code tree structures. The positional dependencies throughout the decoding process are therefore preserved such that the position-wise latent attention can be generalized across diverse code configurations. In this way, the proposed front-padding scheme harmonizes with the polar code construction pattern and enables code-aware attention masking through fixed coordinate indexing for generalization across diverse code lengths. 

\subsection{Embeddings}
We employ three parallel embedding layers that convert the input vectors into $d_{\rm{m}}$-dimensional latent representations ($d_{\rm{m}}\gg N$). In particular, we substitute the standard discrete token embedding with sinusoidal positional encoding in \cite{ref13} with trainable value-aware positional embeddings to capture both value information and positional dependencies in decoding interactions. Due to this architectural modification, the inherent misleading geometric priors in sinusoidal patterns are mitigated during the training. Furthermore, to address the real-valued nature of channel observations, we design a continuous embedding scheme in the Value embedding layer, which eliminates the information degradation caused by quantization in conventional discrete embedding paradigms \cite{ref17.1}.
\subsubsection{Input Embedding}
The input embedding layer projects $\bm{u}^{\rm{p}}$ into ${\mathbf{U}^{\rm{f}}}$, which serves as the input of the LAT decoder. Specifically, ${\mathbf{U}^{\rm{f}}}$ is derived through element-wise multiplication of token value with their corresponding positional embeddings, formulated by
\begin{equation}
\label{input_emb}
{\mathbf{U}^{\rm{f}}}=\bm{u}^{\rm{p}}{\rm{E}^{\rm{in}}_{\rm{p}}}([N_{\rm{max}}]) \text{,}
\end{equation} 
where ${\rm{E}^{\rm{in}}_{\rm{p}}}(\cdot)$ denotes the input positional embedding that maintains $N_{\rm{max}}$ position-specific vectors aligned with the code length. Given that $\bm{u}^{\rm{p}}\in\{0, \pm1\}^{N_{\rm{max}}}$, the resultant ${\mathbf{U}^{\rm{f}}}$ embodies a $d_{\rm{m}}$-dimensional latent representation of frozen bit knowledge constructed by positional basis vectors from ${\rm{E}^{\rm{in}}_{\rm{p}}}([N_{\rm{max}}])$.    

\subsubsection{Value Embedding}
The Value embedding layer transforms $\bm{y}^{\rm{p}}$ into $\mathbf{V}_{\bm{y}}$, serving as a complementary component that augments ${\mathbf{U}^{\rm{f}}}$ to construct the complete Value matrix $\mathbf{V}_{i_{\rm{b}}}$ across all the decoding layers. As a continuous embedding scheme, the Value embedding extends (\ref{input_emb}) through
\begin{equation}
\label{value_emb}
{\mathbf{V}_{\bm{y}}}={\rm{sign}}(\bm{y}^{\rm{p}}){\rm{E}^{\rm{V}}_{\rm{sign}}}([N_{\rm{max}}])+\vert\bm{y}^{\rm{p}}\vert{\rm{E}^{\rm{V}}_{\rm{abs}}}([N_{\rm{max}}]) \text{,}
\end{equation}
where two positional embedding ${\rm{E}^{\rm{V}}_{\rm{sign}}}(\cdot)$ and ${\rm{E}^{\rm{V}}_{\rm{abs}}}(\cdot)$ share the identical architecture with ${\rm{E}^{\rm{in}}_{\rm{p}}}(\cdot)$ and maintain trainable basis vectors to represent the sign and absolute value of $\bm{y}^{\rm{p}}$, respectively. $\vert\bm{y}^{\rm{p}}\vert{\rm{E}^{\rm{V}}_{\rm{abs}}}([N_{\rm{max}}])$ serves as a value-aware representation bias that responds to the real-valued channel observations. Consequently, the quantization-induced information loss is effectively mitigated.

\subsubsection{Query Embedding}
The Query embedding layer integrates two pieces of essential prior knowledge into the Query matrix for all LAT decoding layers: (1) frozen bit constraints and (2) positional dependency bias induced by the frozen bits. To address all possible frozen bit influence patterns on all message bits, this layer implements a position embedding ${\rm{E}^{\rm{Q}}_{\rm{p}}}(\cdot)$ of size $N_{\rm{max}}^2$ to represent position-wise frozen bit knowledge through   
\begin{equation}
\label{Q_emb}
\mathbf{Q}^{\rm{f}}_{i}=\frac{1}{N_{\rm{max}}}\sum_{j\in[N_{\rm{max}}]}{u}^{\rm{p}}_{j}{\rm{E}^{\rm{Q}}_{\rm{p}}}(iN_{\rm{max}}-N_{\rm{max}}+j) \text{,}
\end{equation}
where $\mathbf{Q}^{\rm{f}}_{i}$ represents the $i$-th row of $\mathbf{Q}^{\rm{f}}$. The row-wise reduction of $\bm{u}^{\rm{p}}{\rm{E}^{\rm{Q}}_{\rm{p}}}(\{iN_{\rm{max}}-N_{\rm{max}}+1, \dots, iN_{\rm{max}}\})$ encapsulates the global frozen bit influence on the $i$-th message bit, and thus the output of the Query embedding layer serves as a prior-based bias for the Query matrices across all the decoding layers. Notably, although the frozen bit information is carried in both ${\mathbf{U}^{\rm{f}}}$ and $\mathbf{Q}^{\rm{f}}$, their functional roles diverge fundamentally: $\mathbf{Q}^{\rm{f}}$ strategically drives position-sensitive attention allocation which departs from content-based correlation in conventional self-attention, while ${\mathbf{U}^{\rm{f}}}$ serves as the attention target. 
\subsection{Latent-attention Mechanism \label{LA}}
Typical attention mechanism establishes a dynamic weighting relationship among Value vectors through the compatibility between their pairwise Key vectors and a set of Query vectors. In practice, these vectors are packed into matrices, denoted as $\mathbf{V}$ (Values), $\mathbf{K}$ (Keys), and $\mathbf{Q}$ (Queries), respectively. The proposed latent-attention mechanism retains the fundamental mathematical structure of attention weight computation employed in standard multi-head scaled dot-product attention \cite{ref13}, while $\mathbf{Q}$, $\mathbf{K}$, and $\mathbf{V}$ are uniquely parameterized via layer-specific latent positional encoders (LPE).

\subsubsection{Scaled Dot-product Attention}
The standard scaled dot-product attention function computes the compatibility between $\mathbf{Q}$ and $\mathbf{K}$ through scaled dot-product, formulated as
\begin{equation}
\label{SDattention}
{\rm{Attn}}(\mathbf{Q}, \mathbf{K}, \mathbf{V}) = {\rm{softmax}}\left(\frac{{\mathbf{Q}}{\mathbf{K}^{\rm{T}}}}{\sqrt{d_{\rm{m}}}}\right){\mathbf{V}} \text{,}
\end{equation}
where $\mathbf{K}^{\rm{T}}$ denotes the transpose of $\mathbf{K}$. The softmax function converts the dot-product compatibility into row-wise probabilistic weight distribution and the scaling factor $\sqrt{d_{\rm{m}}}$ prevents the dot-product magnitudes from growing proportionally to $d_{\rm{m}}$, which would otherwise drive the softmax derivatives toward vanishingly small values in high-dimensional latent spaces.

\subsubsection{Multi-head Attention}
By projecting the original $d_{\rm{m}}$-dimensional matrices $\mathbf{Q}$, $\mathbf{K}$, and $\mathbf{V}$ into $h$ parallel $d_{\rm{h}}$-dimensional heads (with $d_{\rm{m}}=h\times d_{\rm{h}}$), the multi-head attention allows simultaneous capture of heterogeneous attention patterns across multiple representation subspaces. Our implementation employs trainable linear layers for both the subspace projections and the final output integration, specifically defined as
\begin{equation}
\label{MHattention1}
{\rm{MultiHead}}(\mathbf{Q}, \mathbf{K}, \mathbf{V}) = {\rm{W}}^{\rm{O}}([{\mathbf{H}}_{1}, \dots, {\mathbf{H}}_{h}]) \text{,}
\end{equation}
where ${\rm{W}}^{\rm{O}}(\cdot)$ denotes the output linear layer, and each head ${\mathbf{H}}_{i}$ is obtained through
\begin{equation}
\label{MHattention2}
{\mathbf{H}}_{i} = {\rm{Attn}}({\rm{W}}^{\rm{Q}}_{i}({\mathbf{Q}}), {\rm{W}}^{\rm{K}}_{i}({\mathbf{K}}),{\rm{W}}^{\rm{V}}_{i}({\mathbf{V}})) \text{,}
\end{equation}
where ${\rm{W}}^{\rm{Q}}_{i}(\cdot)$, ${\rm{W}}^{\rm{K}}_{i}(\cdot)$, and ${\rm{W}}^{\rm{V}}_{i}(\cdot)$ denote the subspace projections on $\mathbf{Q}$, $\mathbf{K}$, and $\mathbf{V}$, respectively. 

\subsubsection{Latent Positional Encoding}
For each decoding layer, we employ a trainable layer-specific LPE for parallel Query encoding and Key encoding, denoted by $\Phi^{\rm{Q}}_{i_{\rm{b}}}: [N_{\rm{max}}] \rightarrow \mathbb{R}^{d_{\rm{m}}}$ and $\Phi^{\rm{K}}_{i_{\rm{b}}}: [2N_{\rm{max}}] \rightarrow \mathbb{R}^{d_{\rm{m}}}$, respectively, for the ${i_{\rm{b}}}$-th decoding layer. The LPEs enable layer-adaptive parameterization of $\mathbf{Q}$ and $\mathbf{K}$ across all decoding layers and therefore facilitate hierarchical attention allocation to capture stepwise polar decoding dependencies among bit positions. Specifically, the parameterization for the ${i_{\rm{b}}}$-th decoding layer is structured as follows:
\begin{equation}
\label{LPE1}
{\mathbf{Q}}_{i_{\rm{b}}} = {\mathbf{Q}^{\rm{f}}}+{\mathbf{P}}_{i_{\rm{b}}} \text{,}
\end{equation}
\begin{equation}
\label{LPE2}
{\mathbf{K}}_{i_{\rm{b}}}=\Phi^{\rm{K}}_{i_{\rm{b}}}([2N_{\rm{max}}]) \text{,}
\end{equation}
\begin{equation}
\label{LPE3}
{\mathbf{V}}_{i_{\rm{b}}}=[{\mathbf{V}}_{\bm{y}},{\mathbf{U}}^{\rm{f}}] \text{,}
\end{equation}
where $\mathbf{P}_{i_{\rm{b}}} = \Phi^{\rm{Q}}_{i_{\rm{b}}}([N_{\rm{max}}])$. The decoding layer inputs ${\mathbf{V}}^{\rm{in}}_{i_{\rm{b}}}$ are derived from decoding layer outputs recursively, initialized as ${\mathbf{V}}^{\rm{in}}_{1}={\mathbf{U}^{\rm{f}}}$. 

% \begin{itemize}
% \item ${\mathbf{Q}}_{i_{\rm{b}}} = {\mathbf{Q}}_{{\bm{u}}^{\rm{f}}}+{\mathbf{P}}_{i_{\rm{b}}}$, where $\mathbf{P}_{i_{\rm{b}}} = \Phi^{\rm{Q}}_{i_{\rm{b}}}([N])$.
% \item ${\mathbf{K}}_{i_{\rm{b}}}=\Phi^{\rm{K}}_{i_{\rm{b}}}([2N])$.
% \item ${\mathbf{V}}_{i_{\rm{b}}}=[{\mathbf{V}}_{\bm{y}},{\mathbf{V}}^{\rm{in}}_{i_{\rm{b}}}]$, where ${\mathbf{V}}^{\rm{in}}_{i_{\rm{b}}}$ derives from decoding layer outputs recursively, initialized as ${\mathbf{V}}^{\rm{in}}_{1}={\mathbf{U}^{\rm{f}}}$.  
% \end{itemize}

Notably, for rate-$1$ (R1) cases (i.e., $R=1$, and ${\mathbf{U}^{\rm{f}}}=\mathbf{0}$), the decoding interactions are simplified to conventional depth first search (DFS) traversal throughout the code tree without prior-based leaf node returns. In these cases, the attention weights are completely dominated by $\Phi^{\rm{Q}}_{i_{\rm{b}}}$ and  $\Phi^{\rm{K}}_{i_{\rm{b}}}$. This architectural property reveals that the LPEs intrinsically learn to model the basic polar decoding dependencies from the pristine code structure. Conversely, for scenarios where $R<1$, the latent-attention activates two complementary mechanisms: (1) The initial concatenate of ${\mathbf{U}^{\rm{f}}}$ for ${\mathbf{V}}_{1}$ to integrate frozen bit values into attention targets, and (2) The additive parameterization of ${\mathbf{Q}}_{i_{\rm{b}}}$ to emulate frozen leaf node returns via trainable Query biases. These latent-attention designs mechanically align with the polar decoding nature by preserving structural dependencies through learned positional patterns while adaptively integrating frozen-bit-induced path modifications. Therefore, the proposed LAT decoder establishes an exact correspondence between the transformer-based DL framework and polar decoding paradigm.

\begin{figure}[!t]
\centering
\includegraphics[width=3.4 in]{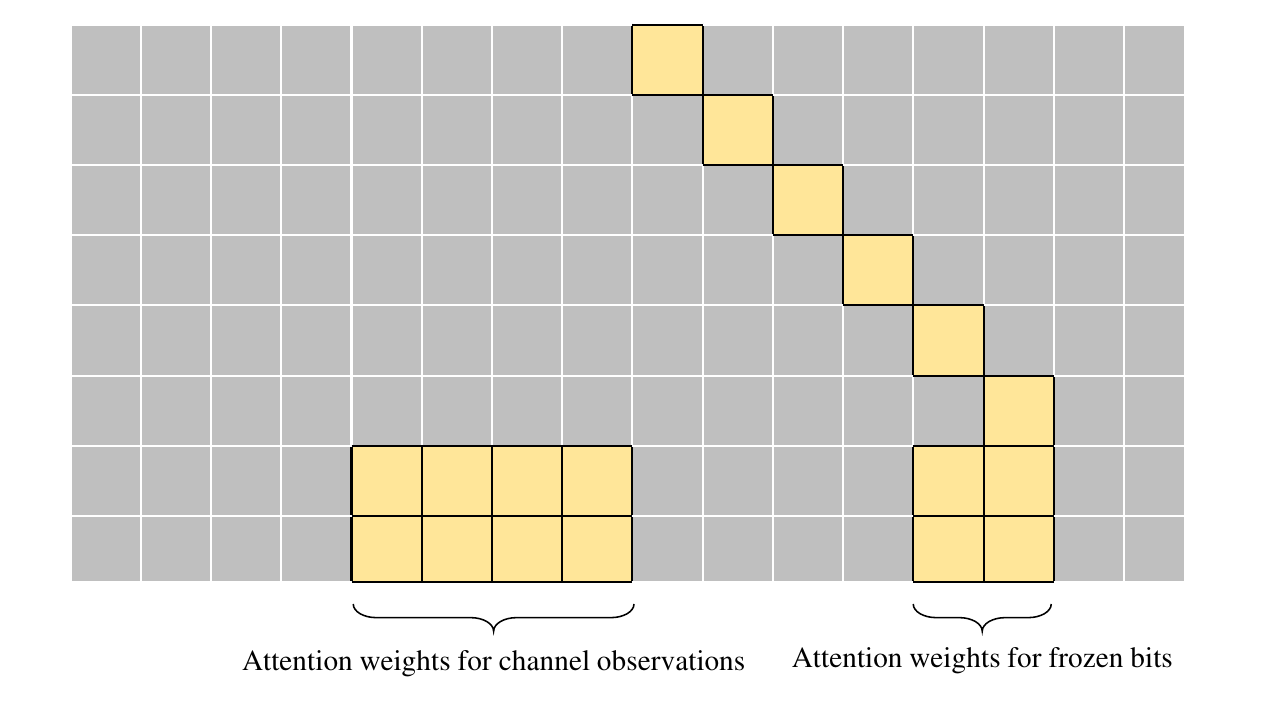}
    \caption{The code-aware masking scheme for $N_{\rm{max}}=8$, $N=4$, and $k=2$. Gray boxes are masked out whereas yellow boxes retain their values. Through this mask, the frozen bits and padding zeros are isolated from other input values while the attention weights over the message rows integrate knowledge from all the channel observations and frozen priors.} 
\label{fig_mask}
\end{figure}

\subsection{Code-aware Mask}
To further extend the latent-attention mechanism mentioned in Section \ref{LA}, we integrate a code-aware masking scheme into the scaled dot-product attention in (\ref{SDattention}) to enforce code-specific attention constraints. Specifically, all values that correspond to illegal connections are set to $-\infty$ at the pre-softmax stage. In practice, we implement a bidirectional masking strategy, where the row-wise masking prevents cross-position attention leakage for both frozen and padding values, while the column-wise masking isolates the padding positions from attention targets. 

\subsubsection{Row-wise Masking}
Let $\mathbf{C}\!=\!\frac{1}{\sqrt{d_{\rm{m}}}}{\mathbf{Q}}{\mathbf{K}^{\rm{T}}}\in\mathbb{R}^{N_{\rm{max}}\times 2N_{\rm{max}}}$ denote the scaled dot-product compatibility. Since frozen bits and front-padding zeros are independent from other input values, the row-wise masking nullifies all non-diagonal positions for both frozen indices and padding rows. Accordingly, the row-wise masked indices are formulated by  
\begin{equation}
\begin{split}
\label{r-masking}
M^{\rm{r}}= \{(i,j)\vert &i\in[N_{\rm{pad}}]\cup\{A_k^{\rm{c}}+N_{\rm{pad}}\},
\\&j \neq i + N_{\rm{max}}\} \text{.}
\end{split}
\end{equation}

\subsubsection{Column-wise Masking}
Since both the front-padding zeros before codewords and non-frozen padding zeros in $\bm{u}^{\rm{f}}$ are independent to the decoding interactions, their corresponding columns are simultaneously nullified for message rows. The column-wise masking set is therefore constructed through
\begin{equation}
\begin{split}
\label{c-masking}
 M^{\rm{c}}= \{(i,j)\vert &i\in\{A_k + N_{\rm{pad}}\}, \\
 &j \in [N_{\rm{pad}}]\cup\{[N_{\rm{pad}}]+N_{\rm{max}}\} \\
 &\cup \{A_k + N_{\rm{pad}}+N_{\rm{max}}\}\} \text{.}
\end{split}
\end{equation}
The entire masked index set is obtained through $M=M^{\rm{r}} \cup M^{\rm{c}}$, and the masked positions of $\mathbf{C}$ are set to $-\infty$, i.e., 
\begin{equation}
\mathbf{C}_{M} = -\infty\text{.} 
\end{equation}
Fig. \ref{fig_mask} demonstrates the code-aware masking scheme for $N_{\rm{max}}=8$, $N=4$, and $k=2$, where gray boxes are masked out while yellow boxes retain their values. Through this masking scheme, the frozen bits and padding zeros are isolated from other input values while the attention weights over the message rows integrate the knowledge from all channel observations and frozen priors. This coordinated masking strategy inherently respects polar coding constraints while maintaining compatibility with variable-length code implementations. In this way, the validity of the attention target is fundamentally guaranteed across diverse coding configurations. Furthermore, since the illegal connections between padding zeros and frozen bits are masked out, the computational complexity of the latent-attention layer is consequently reduced, particularly for short codes and codes with low coding rates.

Notably, in each decoding layer, $\mathbf{C}_{i,j}$ interpretively represents the step-wise positional dependencies between the $i$-th message bit and the $j$-th row of the Value matrix, which is embedded from either the $j$-th padded channel observation for $j\le N_{\rm{max}}$ or the $(j-N_{\rm{max}})$-th padded frozen bit for $j>N_{\rm{max}}$. Therefore, the training of the LAT decoder actually requires abundant training data to cover all possible connections between message bits, received signals and frozen bits. However, since specific positional dependencies are separately preserved, the LAT decoder demonstrates rapid adaptability through test-time fine-tuning.   

% Detailed analysis on the generalization capability of the LAT decoder are given in the Appendix \ref{genralizaiton}. 
\subsection{Typical Transformer-based Sub-architectures}
The remaining parts of the LAT decoder are mainly inherited from typical transformer architectures with minor adjustments to improve training efficiency and stability, including the position-wise feed-forward networks (FFN), the add-and-norm module, and the output module.

\subsubsection{Position-wise FFN}
Beyond the attention modules, each decoding layer incorporates a fully-connected FFN for enhanced feature extraction. These FFNs perform parallel transformation of latent representations across all positions. Each FFN consists of two linear layers and a Mish activation \cite{ref18} in between, with the hidden layer of size $d_{\rm{f}}$. Let $\mathbf{W}_i$ and $\bm{b}_i$ respectively denote the weight matrix and bias vector of the $i$-th linear layer. Each FFN is mathematically formulated as   
\begin{equation}
\label{FFN}
{\rm{FFN}}(\bm{x}) = {\rm{Mish}}(\bm{x}\mathbf{W}_1+\bm{b}_1){\mathbf{W}}_{2} + {\bm{b}}_{2} \text{,}
\end{equation}
where the Mish activation enhances optimization stability through smoother gradient propagation, contrasting with conventional ReLU which employs abrupt zero-boundary. 

\subsubsection{Add-and-norm Module}
Within all the decoding layers, we individually encapsulate the attention modules and FFNs with residual connections, each followed by a layer normalization. These standard add-and-norm modules share identical structure with conventional transformer architectures, formulated as
\begin{equation}
\label{AN}
{\rm{AN}}(\bm{x}) = {\rm{LayerNorm}}({\bm{x}}+{\rm{SubLayer}}(\bm{x})) \text{,}
\end{equation}
where ${\rm{LayerNorm}}(\cdot)$ denotes the layer normalization and ${\rm{SubLayer}}(\cdot)$ denotes either the attention module or FFN. The residual connections facilitate stable gradient flow \cite{ref19}, while the layer normalization ensures position-wise parameter scaling \cite{ref20}.    

\subsubsection{Output Module}
After sequential processing through $N_{\rm{b}}$ decoding layers, the LAT network output $\tilde{\bm{m}}\in\mathbb{R}^{N\times2}$ is obtained through a linear projection and a subsequent softmax probabilistic activation, formulated as
\begin{equation}
\label{out}
\tilde{\bm{m}}={\rm{Output}}(\bm{x}) = {\rm{softmax}}({\bm{x}}\mathbf{W}_{\rm{b}}+{\bm{b}}_{\rm{b}}) \text{,}
\end{equation}
where $\mathbf{W}_{\rm{b}}\in\mathbb{R}^{d_{\rm{m}}\times2}$ and ${\bm{b}}_{\rm{b}}\in\mathbb{R}^2$ denote the weight matrix and bias vector of the output linear layer, respectively. $\tilde{\bm{m}}$ serves as a differentiable bit-wise soft estimation during the training phase, while its positional hard decision $\hat{\bm{m}}\in\{0, 1\}^{N}$, obtained through
\begin{equation}
\hat{m}_{i}=\operatorname*{\arg\max}_{j\in \{0, 1\}} \tilde{m}_{i, j+1} \text{,}
\end{equation}
is adopted for performance evaluation. 

\subsection{Complexity Analysis}
Identical with a typical self-attention mechanism, the latent-attention layer achieves parallel positional processing with constant number ($O(1)$) of sequentially executed operations, whereas a recurrent layer requires $O(N)$ sequential operations and a convolution network with kernel size $k$ requires $O(N/k)$ layers. While the per-layer computational complexity of masked latent-attention reaches $O((2N-k)kd_{\mathrm{m}})$, comparable to conventional self-attention of $O(N^2d_{\mathrm{m}})$ depending on $k$, it remains fundamentally more efficient than recurrent layers of $O(Nd_{\rm{m}}^2)$ and convolution layers of $O(kNd_{\rm{m}}^2)$ for large $d_{\rm{m}}$. This complexity profile preserves the architectural advantage of attention-based models on parallel computation while maintaining practical feasibility for polar decoding tasks.

\section{Methodologies}
Beyond the model structures introduced in Section \ref{LAT}, this section further presents analyses on the methodologies of the proposed LAT decoder, including the mathematical foundation and generalization capability.

\subsection{Mathematical Foundation}
We first outline the mathematical foundation of the LAT decoder. For AWGN channels, the ML decoder minimizes the mean squared error (MSE) between $\bm{y}$ and $\bm{s}$, formulated as
\begin{equation}
\label{ml decoding}
\hat{\bm{m}}_{\rm{ml}}=\operatorname*{\arg\min}_{\bm{x}_{\bm{m}}\in {C}_{\rm{code}}} \Vert \bm{y}-\bm{s}_{\bm{x}}\Vert^2_2 \text{,}
\end{equation}
where $\Vert\cdot\Vert_2$ denotes the Euclidean norm of a vector. Leveraging Tweedie’s formula \cite{ref22}, the optimal estimate of $\bm{m}$ is derived as
\begin{equation}
\label{tweedies formula}
\mathbb{E}\{\bm{m}\vert\bm{y}\}=\mathbb{E}\{\bm{x}\vert\bm{y}\}=\bm{y}+\sigma^2\nabla{\rm{log}}({\rm{p}}(\bm{y})) \text{,}
\end{equation}
where $\nabla{\rm{log}}({\rm{p}}(\bm{y}))$ denotes the score function. In the LAT decoder, the estimator is iteratively implemented through
\begin{equation}
\label{decoding iter}
{\mathbf{V}}_{\rm{in}}^{i_{\rm{b}}+1}={\mathbf{V}}_{\rm{in}}^{i_{\rm{b}}}+{\rm{f}}^{i_{\rm{b}}}({\mathbf{V}}_{\rm{in}}^{i_{\rm{b}}}\vert {\mathbf{V}}_{i_{\rm{b}}}) \text{,}
\end{equation}
where ${\rm{f}}^{i_{\rm{b}}}(\cdot)$ denotes the trainable propagation throughout the $i_{\rm{b}}$-th decoding layer. Since ${\mathbf{V}}_{i_{\rm{b}}}=[{\mathbf{V}}_{\bm{y}},\mathbf{U}^{\rm{f}}]$ is embedded from $\bm{y}$ and $\bm{u}^{\rm{f}}$, which fully characterizes ${\rm{p}}(\bm{y})$, (\ref{decoding iter}) can be rewritten as an approximation of
\begin{equation}
\label{decoding iter2}
{\mathbf{V}}_{\rm{in}}^{i_{\rm{b}}+1}={\mathbf{V}}_{\rm{in}}^{i_{\rm{b}}}+\sigma^2\nabla{\rm{log}}({\rm{p}}( {\mathbf{V}}_{i_{\rm{b}}})) \text{.}
\end{equation}
Then we introduce an energy-based model \cite{ref23}, which constraints ${\rm{p}}(\bm{y})$ via
\begin{equation}
\label{enrgy model}
{\rm{p}}( {\mathbf{V}}_{i_{\rm{b}}})={\rm{C_{e}}exp}(-{\rm{E}}({\mathbf{V}}_{i_{\rm{b}}}\vert{\rm{f}}^{i_{\rm{b}}})) \text{,}
\end{equation}
where ${\rm{C_{e}}}$ denotes an energy constant and the energy function ${\rm{E}}(\cdot)$ is defined as 
\begin{equation}
\label{enrgy function}
{\rm{E}}({\mathbf{V}}\vert{\rm{f}}) = -\left[ {\rm{R}}({\mathbf{V}}) - {\rm{R}_c}({\mathbf{V}}\vert {\rm{f}}) - \lambda \Vert{\mathbf{V}}\Vert_1 \right] \text{,} 
\end{equation}
\begin{equation}
\label{enrgy function2}
{\rm{R}}({\mathbf{V}}) =\frac{1}{2}\log \det\left( \boldsymbol{I} + \gamma \sum_{\mathbf{X}\in\{\mathbf{Q},\mathbf{K},\mathbf{V}\}}  {\mathbf{X}}^{\rm{H}} {\mathbf{X}}^{} \right) \text{,}
\end{equation}
\begin{equation}
\label{enrgy function3}
\begin{split}
{\rm{R}_c}({\mathbf{V}}\vert {\rm{f}}) =& \frac{1}{2} \sum_{i=1}^{h} \log \det\\&
\left( \mathbf{I}
 + \delta\sum_{\mathbf{X}\in\{\mathbf{Q},\mathbf{K},\mathbf{V}\}} (\mathbf{W}_i^{\mathbf{X},\rm{H}} {\mathbf{X}}^{})^{\rm{H}}(\mathbf{W}_i^{\mathbf{X},\rm{H}} {\mathbf{X}}^{})  \right) \text{,}
\end{split}
\end{equation}
where $\Vert\cdot\Vert_1$ denotes the $l1$-norm of a vector, $\lambda$, $\gamma$, and $\delta$ are constants, and $\mathbf{W}_i^{\mathbf{X}}$ denotes the subspace basis matrix for the $i$-th head projecting $\mathbf{X}\in\{\mathbf{Q},\mathbf{K},\mathbf{V}\}$ trained via multi-head attention. For self-attention analysis, where $\mathbf{Q}=\mathbf{K}=\mathbf{V}$, research in \cite{ref24} demonstrates that ${\rm{E}}({\mathbf{V}}\vert{\rm{f}})$ is provably maximizable since ${\rm{R}_c}({\mathbf{V}}\vert {\rm{f}})$ is minimizable through the multi-head attention module while $\lambda \Vert{\mathbf{V}}\Vert_1 - {\rm{R}}({\mathbf{V}})$ is minimizable through the FFNs. Consequently, ${\rm{p}}( {\mathbf{V}}_{i_{\rm{b}}})$ is maximizable through the training of the entire model. Notably, the latent-attention mechanism generates $\mathbf{Q}_{i_{\rm{b}}}$ and $\mathbf{K}_{i_{\rm{b}}}$ via LPEs which eliminates reliance on $\mathbf{V}_{i_{\rm{b}}}$. Therefore, (\ref{enrgy function2}) and (\ref{enrgy function3}) in LAT extend the self-attention mechanism with additive neural approximators, which introduce constant derivative in terms of $\mathbf{V}$. In this way, the training feasibility is maintained while extra knowledge from positional encodings is emphasized. This position-aware architecture aligns the training framework with channel decoding principles by enabling joint optimization of LPEs and standard transformer sublayers without content-based correlations. 

\begin{figure}[!t]
\centering
\includegraphics[width=3.4 in]{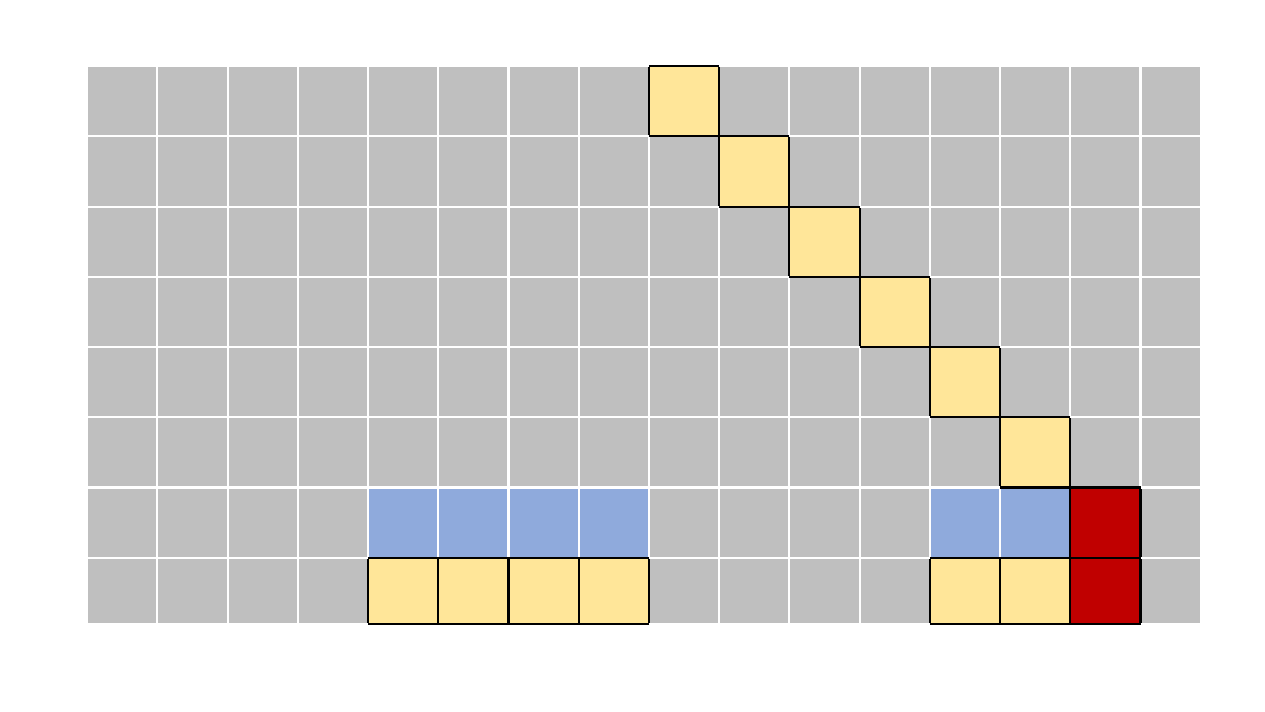}
    \caption{The masking pattern shift from codes with $\{N=4 , k=2\}$ to $\{N=4, k=1\}$. Blue boxes are masked out while red boxes are unmasked. Since there is no overlap between the two masking patterns, the cross-code interference on attention weights is fundamentally eliminated.} 
\label{fig_mask2}
\end{figure}

\subsection{Generalization Analyses}
In this section, we provide analysis on the generalization capability of the LAT decoder across varying code rates and lengths. Front-padding zeros and frozen bits are isolated with other inputs such that for rows $i\in \{A_{\rm{c}}+N_{\rm{pad}}\}\cup[N_{\rm{pad}}]$, the attention outputs retain their input embeddings, ensuring stability for non-message bits. Let $N_{\rm{pad}}^{\rm{f}}=N_{\rm{pad}}+N_{\rm{max}}$ denote the initial index of frozen bit columns in $\mathbf{C}$. For message bits (rows $i\in \{A + N_{\rm{pad}}\}$), the reweighted attention outputs are formulated by
\begin{equation}
\label{generalization}
\mathbf{X}^{(i)}=\sum_{j\in \{[N]+ N_{\rm{pad}}\}}\mathbf{C}_{i, j}\mathbf{V}_{i_{\rm{b}}}^{(j)} +\sum_{j\in \{A_{\rm{c}}+N_{\rm{pad}}^{\rm{f}}\}}\mathbf{C}_{i, j}\mathbf{V}_{i_{\rm{b}}}^{(j)}\text{,}
\end{equation}
where the first term reweights channel observations, and the second term reweights frozen bits. 
\subsubsection{Varying Code Rates}
Changes in $R$ directly alter $A$ and $A_{\rm{c}}$ while these alternations lead to frozen bit pattern shifts. Since the code-aware mask for a frozen bit row $M_i^{\rm{f}}$ and a message bit row $M_i^{\rm{m}}$ are disjoint (i.e., $M_i^{\rm{m}}\cap M_i^{\rm{f}}=\varnothing$), the attention weights remain conflict-free across diverse code rates. Fig. \ref{fig_mask2} demonstrates the masking pattern shift from codes with $\{N=4 , k=2\}$ to $\{N=4, k=1\}$, where the blue boxes are masked out whereas the red boxes are unmasked. Since there is no overlap between two masking patterns, the cross-code interference on attention weights is fundamentally eliminated. Additionally, changes in frozen bit patterns lead to modification of cardinality in $\sum_{j\in \{A_{\rm{c}}+N_{\rm{pad}}^{\rm{f}}\}}\mathbf{C}_{i, j}\mathbf{V}_{i_{\rm{b}}}^{(j)}$, modeling the accumulative contribution of the involved frozen bits, while the prior-based multiplicative influence is adapted through weight bias induced by 
\begin{equation}
\Delta\mathbf{C}_{i,j} =\mathbf{Q}^{\rm{f}}_{i}\mathbf{K}_{i_{\rm{b}}}^{(j)\rm{T}} \text{.}
\end{equation}
Since both of the attention adaptations are configuration-specific and the cross-code interference are eliminated, the LAT decoder is enabled to remain robust to code rate variations. 

\begin{figure}[!t]
\centering
\includegraphics[width=3.4 in]{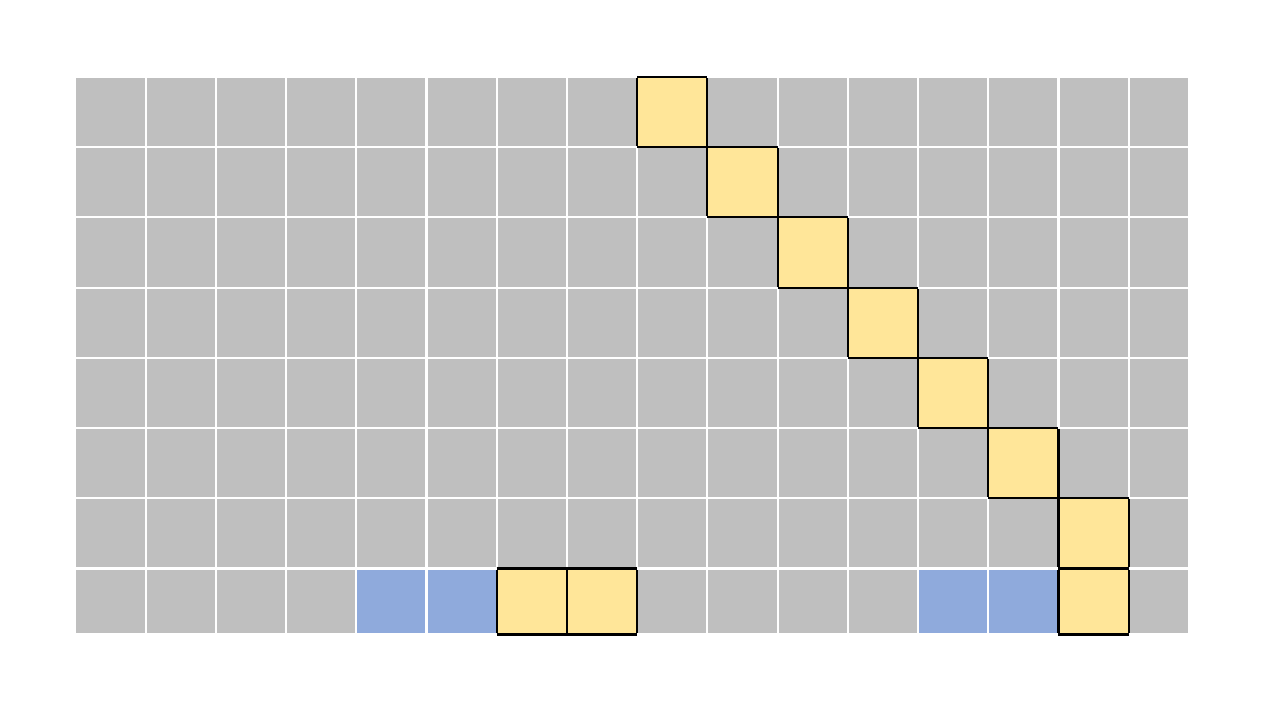}
    \caption{The masking pattern shift from codes with $\{N=4 , k=1\}$ to $\{N=2, k=1\}$. Blue boxes are masked out. Only the rightward subset of the attention weights in the message row are remained, which is consistent with the recursion of the polar code tree.} 
\label{fig_mask3}
\end{figure}

\subsubsection{Varying Code Lengths}
Adjusting $N$ dynamically modifies the masking patterns through a dual mechanism: incremental unmasking of positions during code extension versus progressive masking of value-carrying positions during code contraction. Let $N^{\rm{m}}$, $N^{\rm{m}}_{\rm{pad}}$, $A_{\rm{c}}^{\rm{m}}$, and $N_{\rm{pad}}^{\rm{f,m}}$ parameterize the modified coding configuration. The attention output bias is formulated through 
\begin{align}
% \begin{split}
&\Delta\mathbf{X}^{(i)}_{\rm{f}}=
\sum_{j\in A^{\rm{f,m}}\setminus A^{\rm{f}}}\mathbf{C}_{i, j}\mathbf{V}_{i_{\rm{b}}}^{(j)}-\sum_{j\in A^{\rm{f}}\setminus A^{\rm{f,m}}}\mathbf{C}_{i, j}\mathbf{V}_{i_{\rm{b}}}^{(j)}\text{,}  \\
&\Delta\mathbf{X}^{(i)}_{\bm{y}}=
\begin{cases}
\sum_{j\in \{[N^{\rm{m}}-N]+ N^{\rm{m}}_{\rm{pad}}\}}\mathbf{C}_{i, j}\mathbf{V}_{i_{\rm{b}}}^{(j)}, \quad N^{\rm{m}}>N\\
\!-\!\sum_{j\in \{[N-N^{\rm{m}}]+ N_{\rm{pad}}\}}\mathbf{C}_{i, j}\mathbf{V}_{i_{\rm{b}}}^{(j)}, \text{ otherwise,} 
\end{cases}
% \end{split}
% \end{equation}
\end{align}
where $A^{\rm{f}}=\{A_{\rm{c}}+N_{\rm{pad}}^{\rm{f}}\}$, $A^{\rm{f,m}}=\{A_{\rm{c}}^{\rm{m}}+N_{\rm{pad}}^{\rm{f,m}}\}$, $\Delta\mathbf{X}^{(i)}_{\rm{f}}$ denotes the frozen bit reweighting bias, and $\Delta\mathbf{X}^{(i)}_{\bm{y}}$ denotes channel observation reweighting bias. Notably, the frozen bit interaction adjustment is fundamentally caused by frozen pattern shift, which has been established in prior analyses. Our investigation therefore focuses on the masking shift on channel observation columns. Since the front-padding mechanism preserves codeword bit positional invariance across varying $N$, longer codes extend the attention coverage to the left columns, while shorter codes contract the attention range into rightward column subsets for local optimization. Fig. \ref{fig_mask3} demonstrates the masking pattern shift from codes with $\{N=4 , k=1\}$ to $\{N=2, k=1\}$, where the blue boxes are masked out. Only the rightward subset of the attention weights in the message row are remained, which is consistent with the recursion of the polar code tree. This masking pattern shift exhibits an exact structural alignment with left-iterated polar code trees, where decoding interactions of low-positional bits are complemented through left-induced paths with iterative growth of the code tree, and the right nodes remain recursive consistency in arbitrarily small rightward sub-trees. 

In conclusion, the proposed LAT decoder demonstrates robust compatibility across varying $R$ and $N$ through code-specific decoding interaction adjustments. This investigation reveals an intrinsic system constraint: a critical trade-off exists between universal adaptability and code-specific optimization. Comprehensive training necessitates exhaustive dataset coverage of all valid code configurations to ensure structural generalizability, whereas application-specific implementations benefit from efficient test-time fine-tuning capabilities that enable rapid deployment adaptation.

\section{Training Framework \label{training framework}}
This section delineates the comprehensive training paradigm for the LAT decoder, comprising the dataset preparation, the regularization strategies, and the implementation configurations.

\subsection{Entropy-aware Importance Sampling \label{EIC}}
Within AWGN channels, received signals demonstrate probabilistic clustering around the constellation points, particularly at elevated bit signal-to-noise ratios ($\rm{E_b/N_0}$). This statistical concentration induces rapid decay in occurrence likelihood for distant constellation regions, precisely those critical for characterizing decoding thresholds through model training. To address this sampling-theoretic discrepancy, we propose an entropy-aware importance sampling mechanism that performs uniform sampling of $\bm{n}$ across the 3-$\sigma$ region around $\bm{s}$ depending on the AWGN noise variance and employs an adaptive loss weight $w$ that responds to the noise self-information. By sampling $\bm{n} \sim {\rm{Unif}}([-3\sigma, 3\sigma]^{N})$, we obtain $w$ through  
\begin{equation}
\label{importance sampling2}
 w = 1-\frac{1}{N}{\rm{log}}({\rm{p}}_{\rm{g}}({\bm{n}})) \text{,}
\end{equation}
where  $p_{\rm{g}}(\cdot)$ denotes the probability density function (PDF) of the AWGN. While strategically elevating the sampling frequency in critical low-probability regions, the proposed sampling mechanism emphasizes critical low-probability regions under Gaussian statistical fidelity through entropy-based weighting, effectively improving the decoding threshold characterization without computational overhead.

\subsection{Experience Reflow}
The transient nature of real-time signal sampling during training epochs creates a fundamental tension: while streaming samples enhance spatial coverage for generalization, their ephemeral presence impedes sufficient feature acquisition. To address this learning persistence paradox, let ${\rm{Batch}}_{t}$ denote the original batch of sampled training data at the $t$-th epoch, we propose an experience reflow mechanism as an epoch-wise data augmentation strategy formulated as
\begin{equation}
\begin{split}
\label{ER}
{\rm{Batch}}_{t+1}^{\rm{aug}} =  \{&(\bm{m}^{\star},\bm{y}) \vert (\bm{m},\bm{y})\in   {\rm{Batch}}_{t}^{\rm{aug}}, 
\\ &{\tilde{m}_{i, m_i}}\mathbb{I}({\hat{m}_i} \neq {m}_i)<{\rm{p}}^{\rm{b}},   
\\ &{\hat{\bm{m}}} \neq {\bm{m}}, i\in[N]\}
\cup {\rm{Batch}}_{t+1}^{\rm{sample}}\text{,}
\end{split}
\end{equation}
where ${\rm{Batch}}_{t}^{\rm{aug}}$ denotes the augmented training batch in the $t$-th epoch,   ${\rm{Batch}}_{t}^{\rm{sample}}$ denotes the sampled training batch, ${\rm{p}}^{\rm{b}}$ denotes a confidence threshold for characterizing boundary cases, and $\bm{m}^{\star}$ denotes a reference decoding label obtained through conventional decoders (e.g., SCL decoder or ML decoder). The experience-based data augmentation retains the misclassified samples of a maximum memory buffer size $L_{\rm{m}}$ as a constraint of training complexity and memory cost. Through the experience reflow, samples vulnerable to misclassification (typically near the decoding thresholds) are preferentially conserved, while the progressive accumulation of boundary cases efficiently sharpens the decoding capacity.

\begin{table*}[!t]
\renewcommand\arraystretch{1.5}
\caption{\textcolor{black}{Training Parameters}\label{training para}}
\centering
\begin{tabular}{| c | c | c | c | c | c | c | c |}
\hline
Batch size & $N_{\rm{b}}$ & $d_{\rm{m}}$ & $N_{\rm{max}}$ & $d_{\rm{f}}$ & ${\rm{p}}^{\rm{b}}$ & $L_{\rm{m}}$ & $\lambda_{\rm{B}}$\\
\hline
512 & 6 & 512 & 16 & 2048 & 0.55 & 128 & 0.1\\
\hline
\end{tabular}
\end{table*}

\begin{figure*}[t]
\centering
\subfloat[BER decoding performance.]{
    \centering
    \begin{minipage}[t]{3.4in}
        \centering       
        \includegraphics[width=\linewidth]{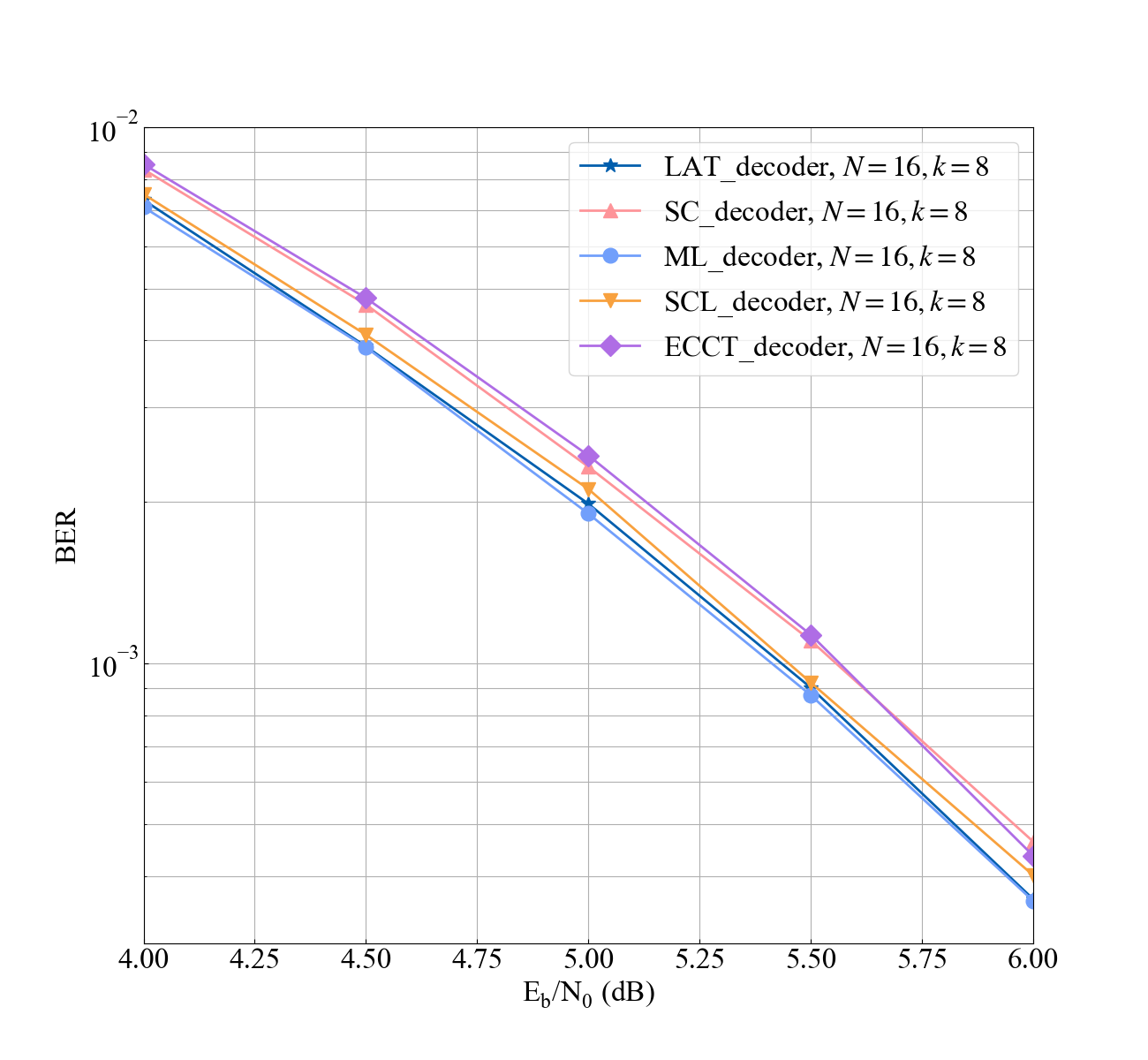}
        \end{minipage}
}%
\subfloat[BLER decoding performance.]{
    \centering
    \begin{minipage}[t]{3.4in}
        \centering       
        \includegraphics[width=\linewidth]{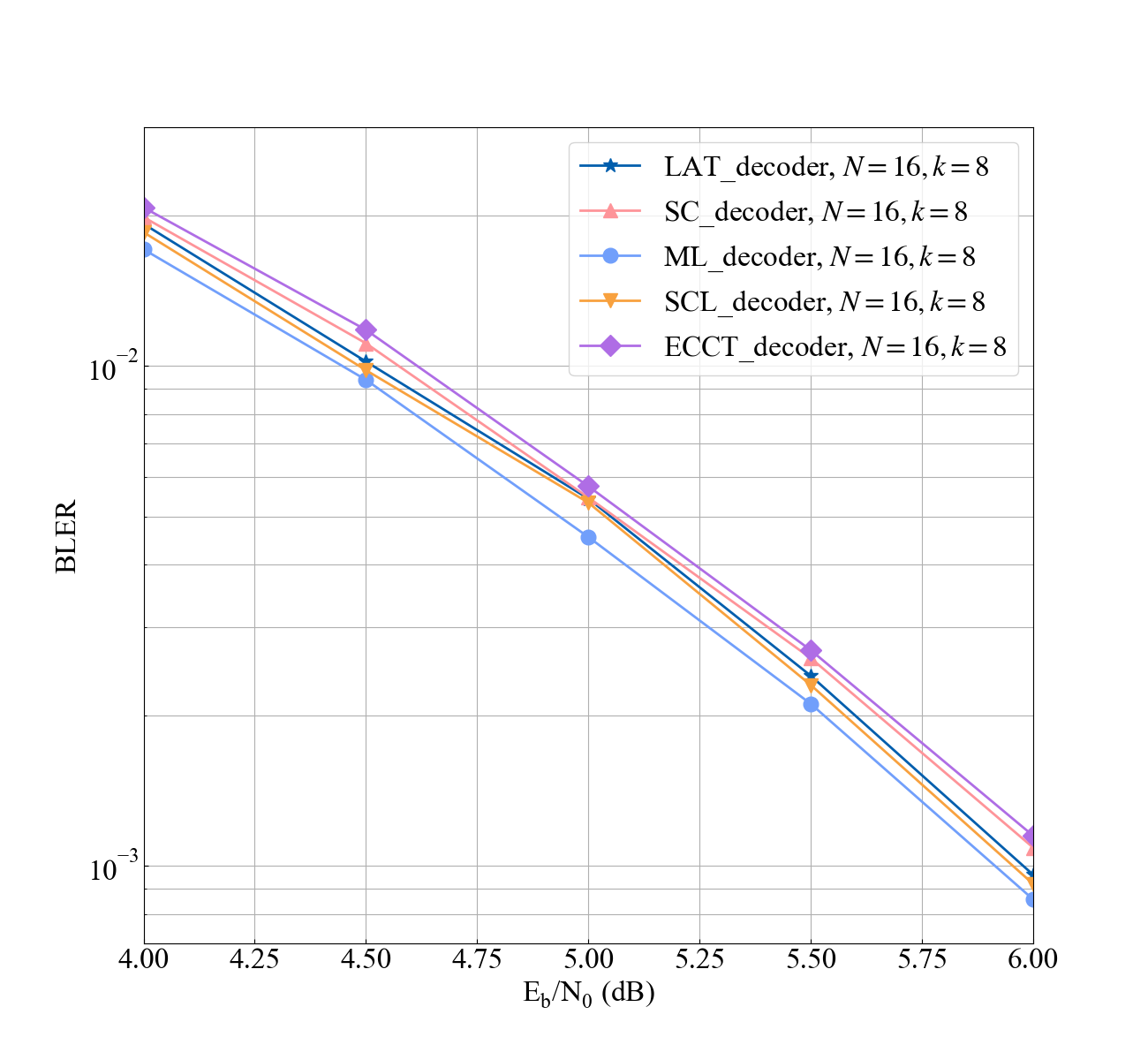}
    \end{minipage}
}%
\caption{\textcolor{black}{Performance of the LAT decoder, the ECCT decoder, the SC decoder, the SCL ($L_{\rm{sc}}=4$) decoder, and the ML decoder with code configuration of $N=16$ and $k=8$, (a) BER, (b) BLER.}}
\label{fig_decoding}
% \vspace{-0.5cm}
\end{figure*} 

\subsection{Label Smoothing}
The binary training labels $\bm{m}^{\rm{b}}\in\{0, 1\}^{N\times2}$ are constructed via positional one-hot encoding of $\bm{m}$, formulated by
\begin{equation}
\label{label}
\bm{m}^{\rm{b}}_{i,j} = \mathbb{I}(\bm{m}_{i}=j-1) \text{,}
\end{equation}
where $j\in[2]$ indices the bit value. Standard label smoothing methods uniformly regularize all samples with fixed smoothing intensity $\epsilon^{\rm{ls}}$ (typically set to $0.1$ for computer vision (CV) tasks \cite{ref21}) through
\begin{equation}
\label{label_smoothing}
\bm{m}^{\rm{ls}}_{i,j} =
\begin{cases}
\epsilon^{\rm{ls}}, & \bm{m}^{\rm{b}}_{i,j}=0 \\
1-\epsilon^{\rm{ls}}, & \bm{m}^{\rm{b}}_{i,j}=1
\text{,}
\end{cases}
\end{equation}
where $\bm{m}^{\rm{ls}}$ denotes the smoothened label. Motivated by the nature of likelihood-based channel decoding, we develop a dynamic label smoothing mechanism where $\epsilon^{\rm{ls}}$ dynamically responds to the noise entropy by
\begin{equation}
\label{Dynamic label_smoothing}
\epsilon^{\rm{ls}} = 0.1{\rm{Tanh}}(w) \text{,}
\end{equation}
where ${\rm{Tanh}}(\cdot)$ denotes a tanh activation that compresses non-negative inputs into $[0, 1)$ for continuous confidence attenuation without over-smoothing. By adaptively aligning label confidence with positional noise self-information, the dynamic label smoothing mechanism efficiently overcomes the rigidity of static smoothing under noisy channels.

\begin{figure*}[t]
\centering
\subfloat[BER decoding performance ($\rm{E_b/N_0}$ variation).]{
    \centering
    \begin{minipage}[t]{3.4in}
        \centering       
        \includegraphics[width=\linewidth]{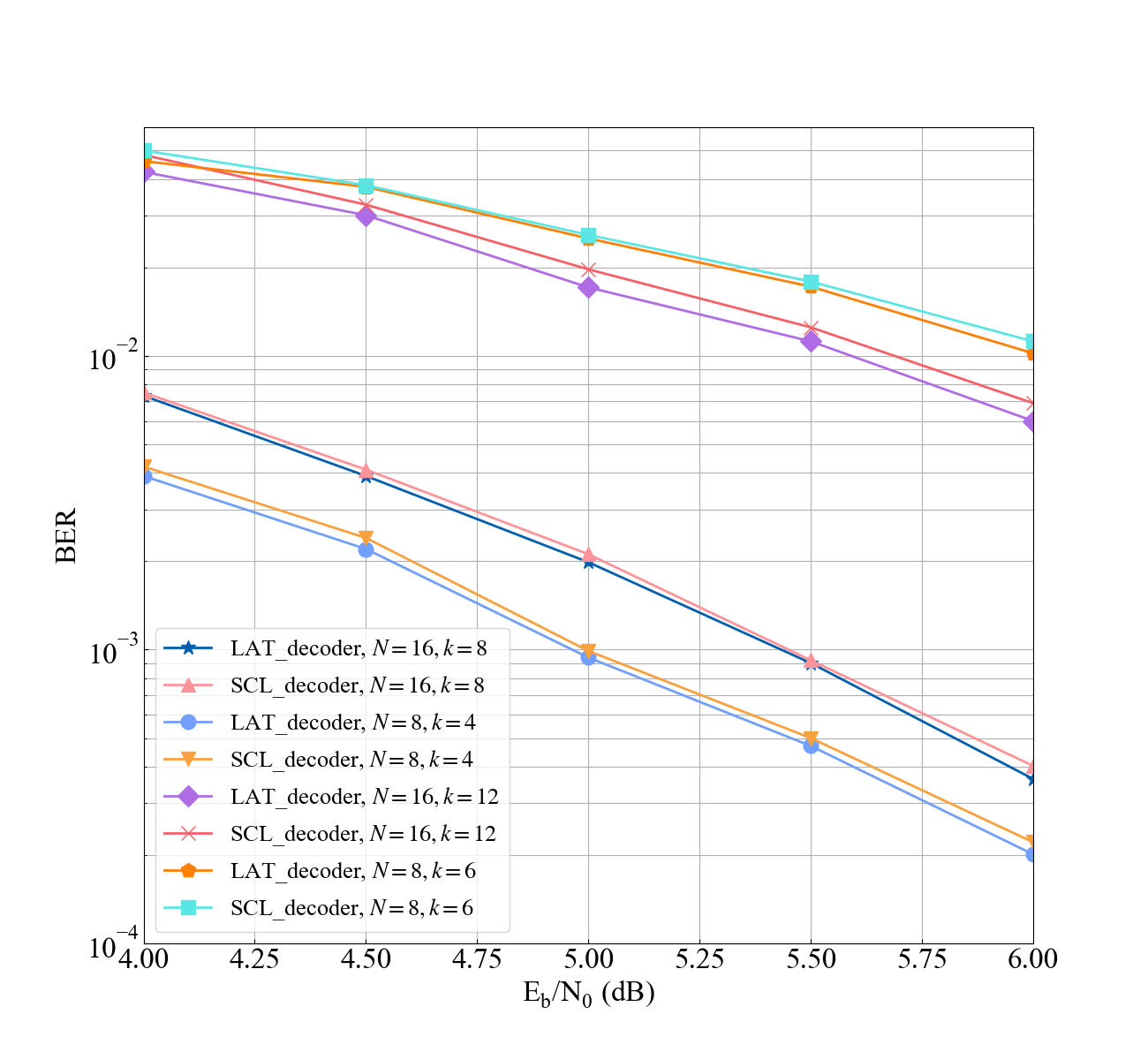}
        \end{minipage}
}%
\subfloat[BLER decoding performance ($\rm{E_b/N_0}$ variation).]{
    \centering
    \begin{minipage}[t]{3.4in}
        \centering       
        \includegraphics[width=\linewidth]{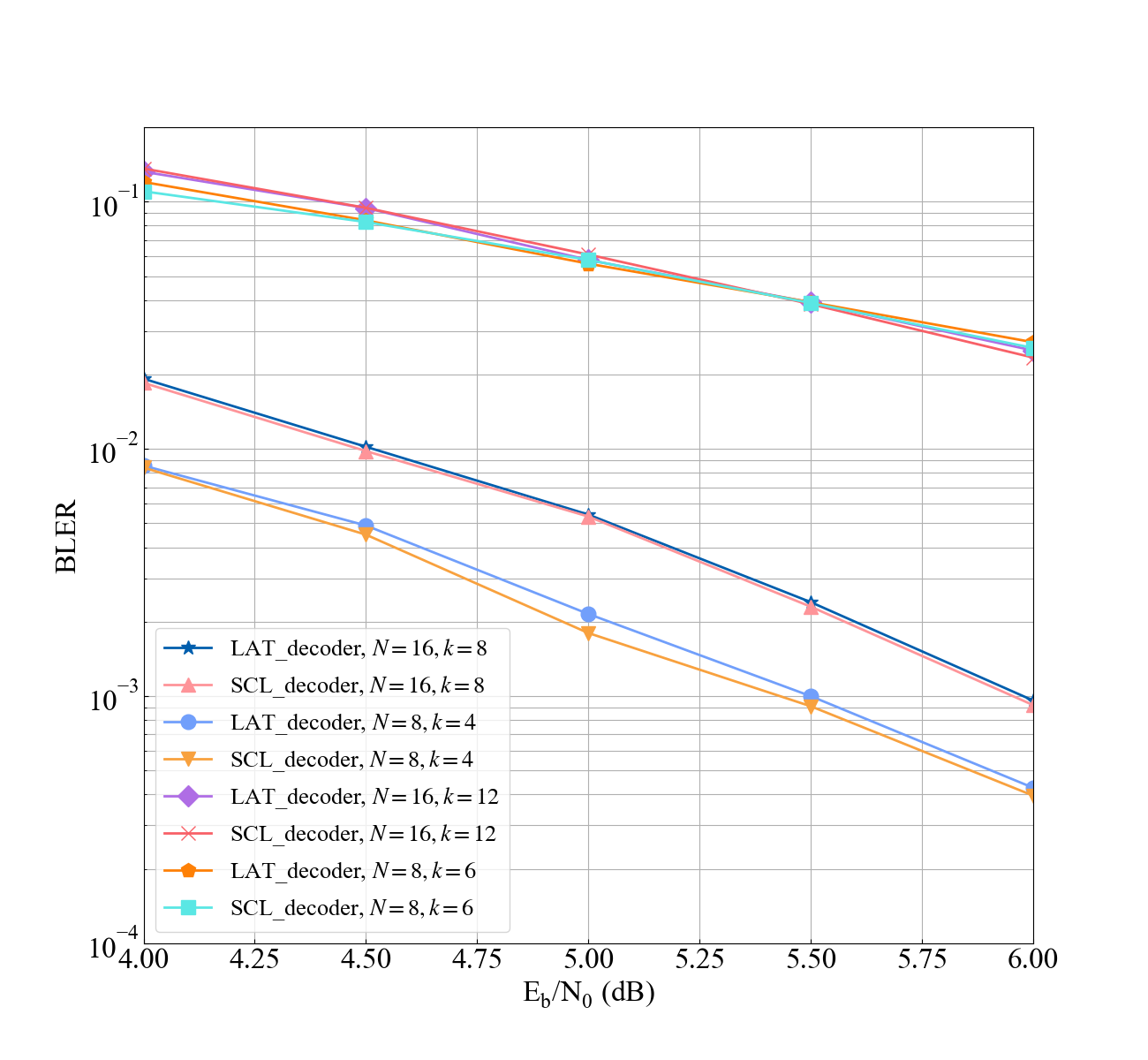}
    \end{minipage}
}\\[1ex]%
\subfloat[BER decoding performance (code rate variation).]{
    \centering
    \begin{minipage}[t]{3.4in}
        \centering       
        \includegraphics[width=\linewidth]{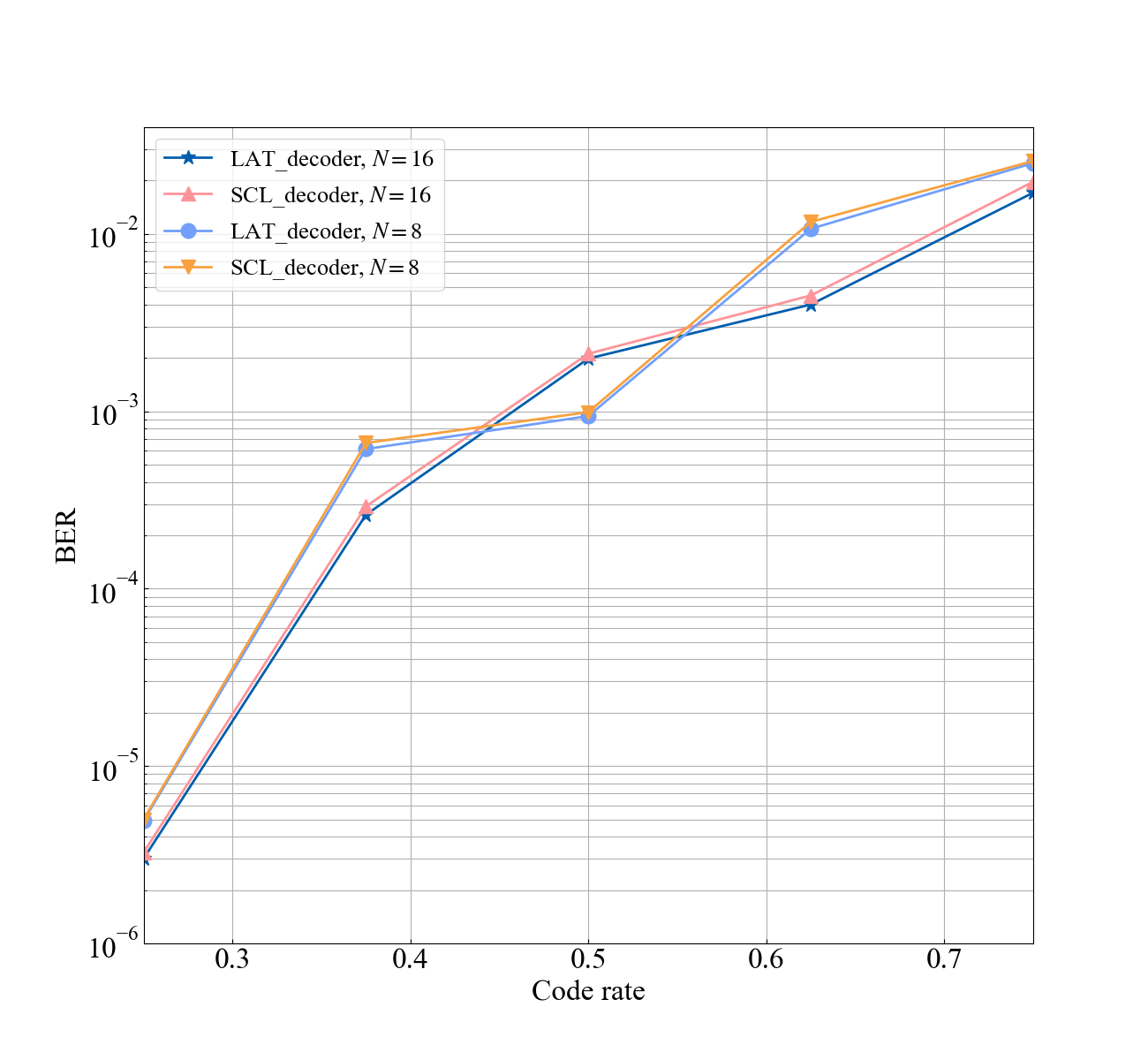}
        \end{minipage}
}%
\subfloat[BLER decoding performance (code rate variation).]{
    \centering
    \begin{minipage}[t]{3.4in}
        \centering       
        \includegraphics[width=\linewidth]{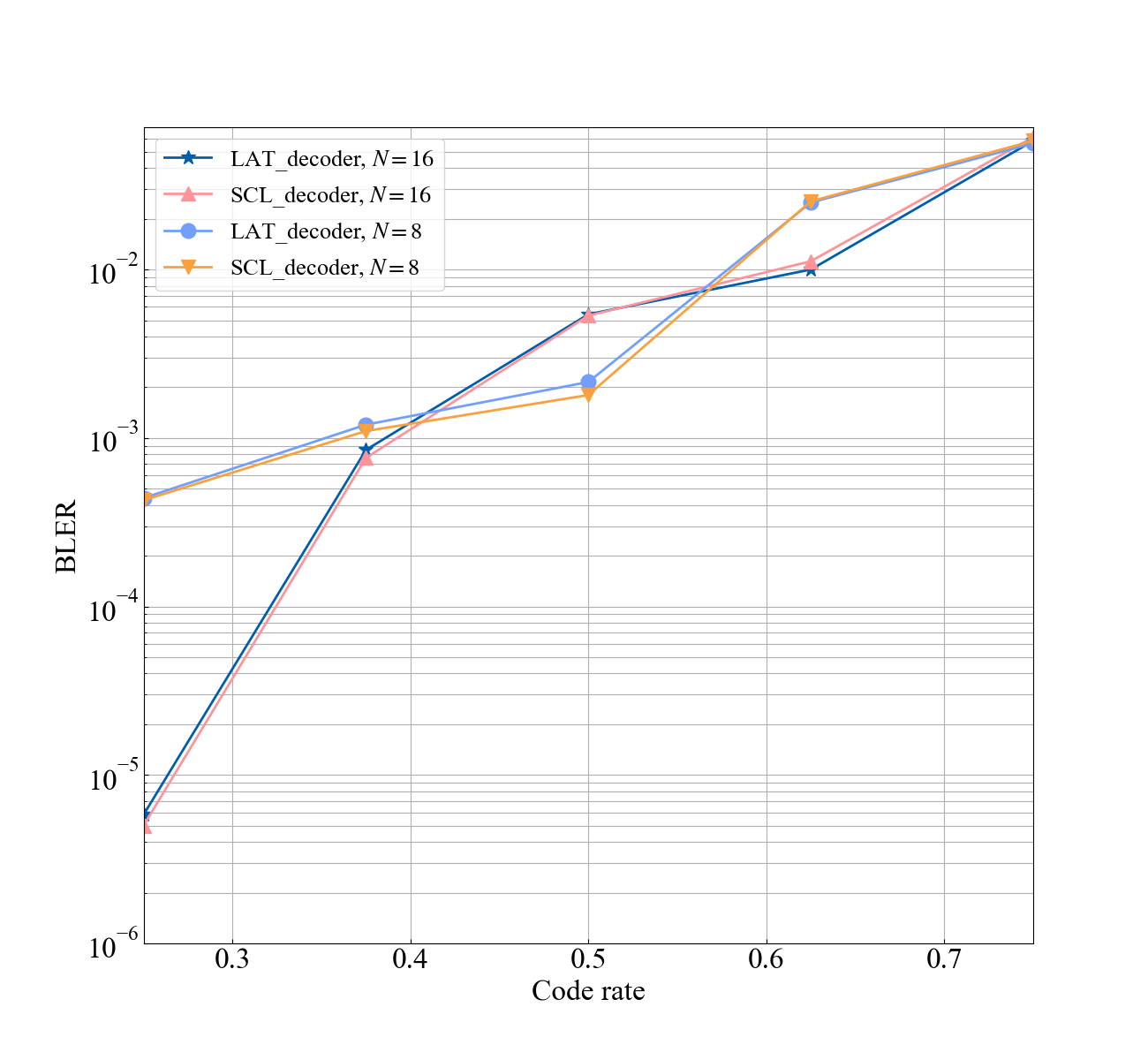}
    \end{minipage}
}%
\caption{\textcolor{black}{Performance of the LAT decoder and the SCL ($L_{\rm{sc}}=4$) decoder, (a) BER and (b) BLER across code configurations of $\{N=16,k=8\}$, $\{N=16,k=12\}$, $\{N=8,k=4\}$, and $\{N=8,k=6\}$; (c) BER and (d) BLER across code rate $R \in [0.25,0.75]$ under $\rm{E}_{\rm{b}}/\rm{N}_{\rm{0}} = 5~\mathrm{dB}$ for $N\in\{8, 16\}$.}}
\label{fig_decoding2}
% \vspace{-0.5cm}
\end{figure*}

\begin{figure*}[t]
\centering
\subfloat[BER decoding performance.]{
    \centering
    \begin{minipage}[t]{3.4in}
        \centering       
        \includegraphics[width=\linewidth]{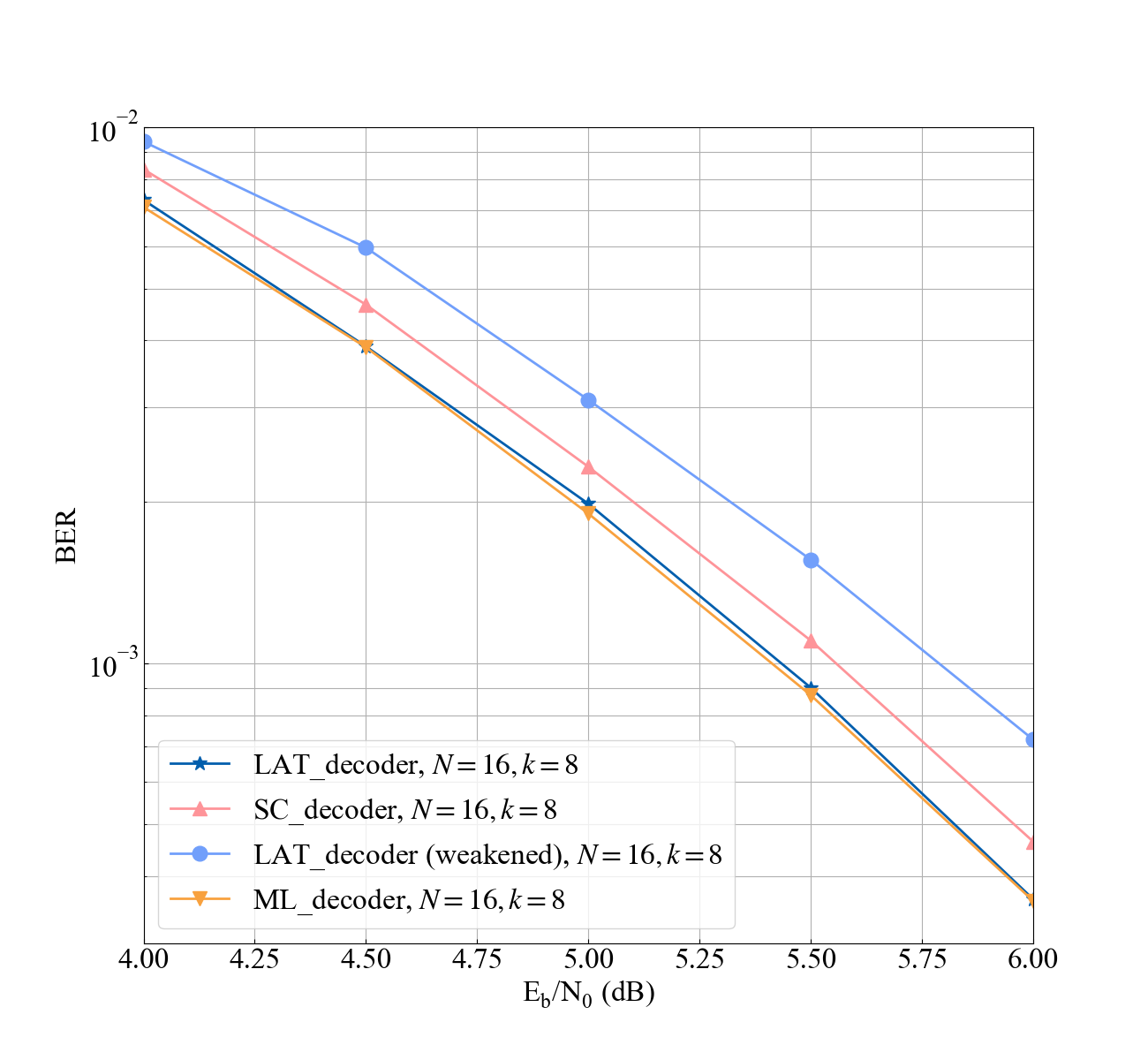}
        \end{minipage}
}%
\subfloat[BLER decoding performance.]{
    \centering
    \begin{minipage}[t]{3.4in}
        \centering       
        \includegraphics[width=\linewidth]{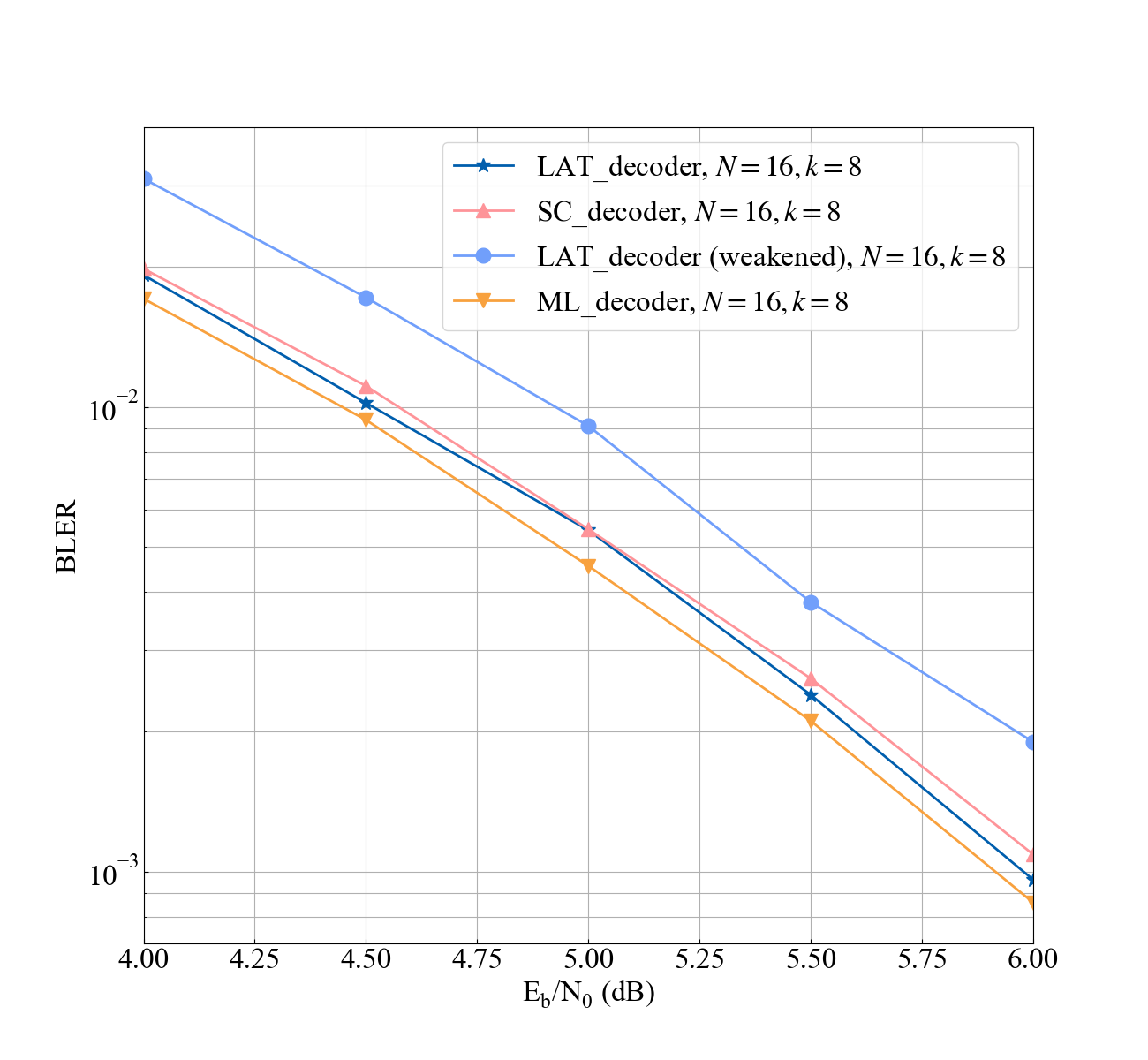}
    \end{minipage}
}%
\caption{\textcolor{black}{Performance of the LAT decoder, and the weakened LAT decoder with code configurations of $N=16$ and $k=8$, (a) BER, (b) BLER.}}

\label{fig_decoding3}
% \vspace{-0.5cm}
\end{figure*}

\subsection{Loss Function}
Diverging from conventional binary cross-entropy (BCE) decoding loss, since the dynamic label smoothing softens the decoding labels, we use a weighted Kullback-Leibler divergence (KLD) for the continuous optimization, formulated by
\begin{equation}
\label{KLD}
{\mathcal{L}_{\rm{KLD}}}({\tilde{\bm{m}}}, {\bm{m}}^{\rm{ls}}) = \frac{w}{2N}\sum_{i,j} {\bm{m}}^{\rm{ls}}_{i,j}{\rm{log}}\frac{{\bm{m}}^{\rm{ls}}_{i,j}}{{\tilde{{\bm{m}}}_{i,j}}} \text{.}
\end{equation}
Notably, while KLD minimizes bit-wise probability distribution mismatches, it fails to directly optimize the BLER. To reconcile this discrepancy between continuous optimization objective and discrete performance metric, we extend the KLD loss by constructing a multi-objective learning framework through loss function hybridization by 
\begin{equation}
\label{Loss}
{\mathcal{L}} = {\mathcal{L}_{\rm{KLD}}} + \lambda_{\rm{B}}{\mathcal{L}_{B}}, \quad \mathcal{L}_{B}=\mathbb{E}^{\rm{train}}(\mathbb{I}(\hat{\bm{m}}={\bm{m}})) \text{,}
\end{equation}
where the hyperparameter $\lambda_{\rm{B}}$ denotes the weighting coefficient and $\mathcal{L}_{B}$ injects BLER awareness over the training dataset as a regularizer. This synergistic integration enables joint optimization of probabilistic fidelity and discrete decoding performance, effectively bridging the gap between training dynamics and testing metrics.

\subsection{Implementation Details}
We use Adam as the optimizer with the learning rate of 0.0002 and exponential decay rate of (0.9, 0.98), integrated with an $l2$ regularizer for training weight decay. The model training is implemented on a 24 GB NVIDIA RTX 4090 and 4 NVIDIA A100-PCIE-40GB GPUs, with the parameters listed in Table \ref{training para}. The training dataset is obtained through entropy-aware importance sampling in Section \ref{EIC} while the validation dataset and testing dataset are sampled from AWGN channels. We also employ an early stopping strategy that terminates the training for a staged patience of 100 epochs without validation performance improvement and a maximum patience of 2000 total epochs. 

\section{Numerical Results}
\subsection{Simulation Setup \label{setup}}
To evaluate the performance of the proposed LAT decoder, we trained our model on standard short-length polar decoding scenarios under AWGN channels. The frozen positions are set through GA-based LLR analysis \cite{ref17}, while the code configurations are uniformly sampled from all valid settings (i.e., $k<N=2^{n}<N_{\rm{max}}$). The $\rm{E_b/N_0}$ (dB) of channel is sampled from ${\rm{Unif}([15])}$ for each training batch. Besides the LAT decoder, we also setup a typical SC detector, a typical SCL decoder with list size $L_{\rm{sc}}=4$, a typical ECCT \cite{ref14} decoder with $N^{\rm{E}}=6$ decoding layers, and an ML detector for comparison. Our experiments are conducted under conda 23.1.0, python 3.9.16, and pytorch 2.4.0, where python is the coding language, conda is the environment for DL implementations, and pytorch is a common framework to develop DL models.  

 % We separately train two LAT decoders parameterized with $d_{\rm{m}}\in\{512, 1024\}$, $N_{\rm{max}}\in\{16, 32\}$, and $d_{\rm{f}}\in\{1024, 2048\}$, respectively for performance evaluation,

\subsection{Decoding Performance Against Benchmarks \label{exp1}}
Firstly, we evaluate the BER and BLER performance of the five decoders with code configurations of $N=16$ and $k=8$. The message sequences are uniformly sampled from $\{0, 1\}^{N}$ and the frozen bits are set to $0$. The codewords are obtained through (\ref{polar_encode3}) without reverse shuffle matrices, consistent with the polar code tree. A standard BPSK modulator is employed to generate the transmit signal through $\bm{s}=1-2\bm{x}$. With $\rm E_b/N_0$ from 4 dB to 6 dB, We record the average BER and BLER of the decoders respectively for comparison as an assessment. The LAT decoder is trained for additional 100 epochs as fine-tuning for a specific code configuration before test.

% The decoding performance of the LAT decoder is compared against other benchmarks 

Experiments demonstrate that for $\rm E_b/N_0$ from 4 dB to 6 dB, as illustrated in Fig. \ref{fig_decoding}, the BER and BLER of the LAT decoder outperforms the SC decoder by an average margin of 0.80 dB in BER and 0.28 dB in BLER, and achieves comparable performance to the ML and SCL decoder. Notably, the LAT decoder achieves the nearest BER performance to the ML decoder compared to other benchmarks. The ECCT decoder is outperformed by the LAT decoder in both the BER and BLER, which demonstrates the benefit of the propose latent attention mechanism over conventional self-attention. The more pronounced BER advantage over BLER improvement primarily stems from the dominant role of the continuous KLD optimization in the objective function. Although the BLER-aware regularization introduced by $\mathcal{L}_{B}$ enhances BLER performance, its inherent non-differentiability and discontinuity fundamentally constrain its optimization potential, resulting in relatively modest BLER gains alongside training instability.

\subsection{\textcolor{black}{Adaptability Across Diverse Code Configurations} \label{exp2}}
\textcolor{black}{To evaluate the generalization capability of the LAT decoder, in this section we test its BER and BLER performance across diverse code configurations. Extending the experimental framework established in Section \ref{exp1} for $\rm{E_b/N_0}$ variation analysis, we conduct additional investigations under fixed $\rm{E}_{\rm{b}}/\rm{N}_{\rm{0}}$ of $5$ dB with code rates $R \in [0.25, 0.75]$ for $N \in \{8,16\}$. The SCL decoder serves as a near ML benchmark. The ECCT decoder is excluded from comparison since its code-fixed architecture requires dedicated network settings for each code specification and necessitates complete retraining for new code configurations. In contrast, the LAT decoder achieves configuration adaptation through quick targeted fine-tuning and maintains architectural consistency across varying code parameters without structural reconfiguration. As demonstrated in Fig. \ref{fig_decoding2}, the LAT decoder maintains configuration-robust performance comparable to SCL decoder across all tested codes. For codes with $N=8$ and $k=4$, in particular, the decoding performance of the LAT decoder is close to the SCL decoder in BER and slightly outperformed by SCL decoder in BLER of an average gap of 0.39 dB. However, for codes with larger length and higher rates, the performance gap is narrowed and even surpassed, revealing stable generalization capabilities of the LAT decoder. Notably, performance degradation occurs when no fine-tuning is employed before testing. Nevertheless, the LAT decoder exhibits rapid adaptability through test-time fine-tuning, thereby establishing a novel deep learning paradigm of universal trainable decoders compared to conventional configuration-fixed DL models.}

\subsection{Benefit of the Modified Training Strategy \label{exp3}}
Since error labels are inevitably introduced into the training dataset through sampling over noise, the effectiveness of the training strategy plays a critical role in determining model performance. To further evaluate the benefit of the proposed training strategy, we parallelly train a weakened version of LAT decoder with identical initialization, training epochs, and testing configurations as the proposed LAT decoder in Section \ref{exp1} whereas under a simplified training framework without importance sampling and label smoothing mentioned in Section \ref{training framework}. Although the simplified training protocol enables sampling from low-probability regions through AWGN channels with low $\rm E_b/N_0$, this sampling approach simultaneously introduces significant label noise that substantially constraints the decoding performance. Therefore, the experience reflow is preserved for necessary empirical labels to characterize the decoding boundaries. The decoding performance of the weakened LAT decoder is compared with the proposed LAT decoder, SC decoder and ML decoder as an assessment on the training framework. Fig. \ref{fig_decoding3} illustrates that the weakened LAT decoder suffers serious performance degradation of 1.24 dB in BER and 1.59 dB in BLER against the SC decoder and the gap is even larger compared to the proposed LAT decoder, revealing that the proposed training framework makes critical contributions to improve the decoding capacity through emphasized low-probability regions and smoothened labels.     

\section{Conclusion}
This paper proposes a latent-attention based transformer decoder specifically designed for short-length polar codes, which achieves near ML performance through synergistic integration of code structure awareness and neural attention mechanisms. Through value-aware positional embeddings, the channel observations and frozen bit knowledge are encoded into position-wise latent representations, used for constructing the attention patterns across all decoding layers. In addition, each decoding layer employs a specific latent positional encoder to augment the attention patterns with layer-wise positional dependencies. \textcolor{black}{While a front padding scheme is employed to preserve the inherent polar decoding positional dependencies across diverse code configurations by retaining the absolute message bit indices, a code-aware masking strategy is used accordingly to isolate illegal attention connections and enable cross-code adaptability.} The training framework combines three pivotal components: an entropy-aware importance sampling strategy that emphasizes low-probability error-prone regions, an experience reflow mechanism that stabilizes decoding thresholds through dynamic boundary characterization, and a dynamic label smoothing that balances model confidence with likelihood-based channel decoding nature. Additionally, the integration of a BLER-aware regularization term within the KLD loss function bridges probabilistic modeling fidelity with discrete decoding performance metrics. Experimental validation demonstrates that the LAT decoder achieves near ML performance for short codes  while maintaining generalization capabilities across diverse code configurations. Compare to conventional configuration-fixed DL-based models, the general trainable LAT decoder indicates a versatile solution for next-generation communication systems, establishing a new paradigm for neural decoding that harmonizes algebraic code structures with DL-based methodologies.           

{\appendices}

% Considering two different codes with code lengths of $N_1, N_2$ and message length of $k_1, k_2$, respectively. Then the padding length of them are $N_{\rm{pad}}^{1}$ and $N_{\rm{pad}}^{2}$, respectively. 

\vfill

\end{document}